\documentclass[oldversion]{aa}
\usepackage{graphicx}
\def\H0 {$H_{\rm o}$}

\def\CH3C2H {\hbox{${\rm CH}_3{\rm C}_2{\rm H}$}} %CH3C2H
\def \la{\mathrel{\mathchoice   {\vcenter{\offinterlineskip\halign{\hfil
$\displaystyle##$\hfil\cr<\cr\sim\cr}}}
{\vcenter{\offinterlineskip\halign{\hfil$\textstyle##$\hfil\cr
<\cr\sim\cr}}}
{\vcenter{\offinterlineskip\halign{\hfil$\scriptstyle##$\hfil\cr
<\cr\sim\cr}}}
{\vcenter{\offinterlineskip\halign{\hfil$\scriptscriptstyle##$\hfil\cr
<\cr\sim\cr}}}}}
\def \ga{\mathrel{\mathchoice   {\vcenter{\offinterlineskip\halign{\hfil
$\displaystyle##$\hfil\cr>\cr\sim\cr}}}
{\vcenter{\offinterlineskip\halign{\hfil$\textstyle##$\hfil\cr
>\cr\sim\cr}}}
{\vcenter{\offinterlineskip\halign{\hfil$\scriptstyle##$\hfil\cr
>\cr\sim\cr}}}
{\vcenter{\offinterlineskip\halign{\hfil$\scriptscriptstyle##$\hfil\cr
>\cr\sim\cr}}}}}

\begin{document}

\title{On carbon and oxygen isotope ratios in starburst galaxies: New data from
       NGC~253 and Mrk~231 and their implications}

\author{C. Henkel\inst{1,2} 
        \and 
        H. Asiri\inst{2}
        \and
        Y. Ao\inst{1,3,4}
        \and
        S. Aalto\inst{5}
        \and
        A.~L.~R. Danielson\inst{6}
        \and
        P.~P. Papadopoulos\inst{7,8}
        \and
        S. Garc\'{\i}a-Burillo\inst{9}
        \and
        R. Aladro\inst{10}
        \and
        C.~M.~V. Impellizzeri\inst{10}
        \and
        R. Mauersberger\inst{10}
        \and
        S. Mart\'{\i}n\inst{11}
        \and 
        N. Harada\inst{1}}

\offprints{C. Henkel, \email{chenkel@mpifr-bonn.mpg.de}}

\institute{
  Max-Planck-Institut f{\"u}r Radioastronomie, Auf dem H{\"u}gel 69, 53121 Bonn, Germany
 \and 
  Astronomy Department, Faculty of Science, King Abdulaziz University, P.O. Box 80203, Jeddah 21589, 
  Saudi Arabia
 \and
  Purple Mountain Observatory, Chinese Academy of Sciences, 21008, Nanjing, China
 \and
  National Astronomical Observatory of Japan, 2--21--1 Osawa, Mitaka, Tokyo 181--8588, Japan
 \and
  Dept. of Earth and Space Sciences, Chalmers University of Technology, Onsala Observatory, 
  43994 Onsala, Sweden
 \and 
  Institute for Computational Cosmology, Dept. of Physics, Durham University, South Road,
  Durham DH1 3LE, UK
 \and
  School of Physics, Cardiff University, Queens Building, The Parade, Cardiff CF24 3AA, UK
 \and
  European Southern Observatory, Karl-Schwarzschild-Strasse 2, 85748 Garching, Germany
 \and
  Observatorio Astron{\'o}mico Nacional (OAN) - Observatorio de Madrid, Alfonso XII, 3, 28014 Madrid, Spain
 \and
  European Southern Observatory, Avda. Alonso de C{\'o}rdova 3107, Vitacura, Casilla 19001, Santiago, Chile
 \and
  Institute de Radioastronomie Millim{\'e}trique, Rue de la Piscine 300, 38406 Saint-Martin d'H{\`e}res, France
}
 
\date{Received date ; accepted date}
 
\abstract
{Carbon and oxygen isotope ratios are excellent measures of nuclear processing, but few such data 
have been taken toward extragalactic targets so far. Therefore, using the IRAM 30-m telescope, 
CN and CO isotopologues have been measured toward the nearby starburst galaxy NGC~253 and the 
prototypical ultraluminous infrared galaxy Mrk~231. Toward the center of NGC~253, the CN and 
$^{13}$CN $N$ = 1$\rightarrow$0 lines indicate no significant deviations from expected local 
thermodynamical equilibrium after accounting for moderate saturation effects (10 and 25\%) in 
the two detected spectral components of the main species. Also accounting for calibration 
uncertainties, which dominate the error budget, the $^{12}$C/$^{13}$C ratio becomes 40$\pm$10. 
This is larger than the ratio in the central molecular zone of the Galaxy, suggesting a higher 
infall rate of poorly processed gas toward the central region. Assuming that the ratio also holds 
for the CO emitting gas, this yields $^{16}$O/$^{18}$O = 145$\pm$36 and $^{16}$O/$^{17}$O = 
1290$\pm$365 and a $^{32}$S/$^{34}$S ratio close to that measured for the local interstellar 
medium (20--25). No indication for vibrationally excited CN is found in the lower frequency fine 
structure components of the $N$ = 1$\rightarrow$0 and 2$\rightarrow$1 transitions at rms noise 
levels of 3 and 4\,mK (15 and 20\,mJy) in 8.5\,km\,s$^{-1}$ wide channels. Peak line intensity
ratios between NGC~253 and Mrk~231 are $\sim$100 for $^{12}$C$^{16}$O and $^{12}$C$^{18}$O 
$J$ = 1$\rightarrow$0, while the ratio for $^{13}$C$^{16}$O $J$ = 1$\rightarrow$0 is $\sim$250. 
This and similar $^{13}$CO and C$^{18}$O line intensities in the $J$ = 1$\rightarrow$0 and 
2$\rightarrow$1 transitions of Mrk~231 suggest $^{12}$C/$^{13}$C $\sim$ 100 and $^{16}$O/$^{18}$O 
$\sim$ 100, in agreement with values obtained for the less evolved ultraluminous merger Arp~220. 
Also accounting for other (scarcely available) extragalactic data, $^{12}$C/$^{13}$C ratios appear 
to vary over a full order of magnitude, from $>$100 in ultraluminous high redshift galaxies to 
$\sim$100 in more local such galaxies to $\sim$40 in weaker starbursts not undergoing a large 
scale merger to 25 in the Central Molecular Zone of the Milky Way. With $^{12}$C being predominantly 
synthesized in massive stars, while $^{13}$C is mostly ejected by longer lived lower mass stars at 
later times, this is qualitatively consistent with our results of decreasing carbon isotope ratios with 
time and rising metallicity. It is emphasized, however, that both infall of poorly processed 
material, initiating a nuclear starburst, as well as the ejecta from newly formed massive stars  
(in particular in case of a top-heavy stellar initial mass function) can raise the carbon isotope 
ratio for a limited amount of time.}


\titlerunning{$^{12}$C/$^{13}$C in NGC~253}

\authorrunning{Henkel, C., Asiri, H., Ao, Y. et al.}

\maketitle

\section{Introduction}

When studying stellar nucleosynthesis and chemical enrichment, it is 
difficult to optically distinguish between isotopes of a given element, 
since their atomic lines are blended. However, microwave lines from rare
isotopic substitutions of a given molecular species, so-called 
``isotopologues'', are well separated from their parent molecule, typically 
by a few percent of their rest frequency. Thus, the frequencies of the main
and rare species are close enough to be observed with the same technical 
equipment but without the problem of blending.

A few years ago, it became apparent (Wouterloot et al. 2008; Wang et al. 
2009) that with respect to its composition, the metal poor outer 
Galaxy does not provide a ``bridge'' between the solar neighborhood 
and the even more metal poor Large Magellanic Cloud (LMC). This 
can be explained by the different age of the bulk of the stellar 
populations of the outer Galaxy and the LMC and can be exemplified 
by one of the most thoroughly studied isotope ratios, that of carbon. 
The two stable isotopes, $^{12}$C and $^{13}$C, have been measured 
throughout the Galaxy, in prominent star forming regions of the LMC, 
and in a large number of stellar objects (e.g., Milam et al. 2005; Wang
et al. 2009; Abia et al. 2012; Mikolaitis et al. 2012). The $^{12}$C/$^{13}$C 
ratio is a measure of ``primary'' versus ``secondary'' processing. 
$^{12}$C is produced on rapid timescales primarily via He burning in 
massive stars. $^{13}$C is mainly produced via CNO processing of 
$^{12}$C seeds from earlier stellar generations. This occurs on a 
slower time scale during the red giant phase in low and intermediate 
mass stars or novae (for reviews, see Henkel et al. 1994; Wilson \& 
Rood 1994).

Previous observations (e.g., Henkel et al. 1985; Stahl et al.
1989; Wouterloot \& Brand 1996; Milam et al. 2005; Sheffer et al.
2007) have demonstrated that the $^{12}$C/$^{13}$C ratio can vary 
strongly within the Galaxy. In the outer Galaxy very high ratios
of $^{12}$C/$^{13}$C $>$100 are found; in the local interstellar 
medium $^{12}$C/$^{13}$C $\sim$ 70, while in the inner Galactic 
disk and Large Magellanic Cloud (LMC) $^{12}$C/$^{13}$C $\sim$ 50. 
The solar system ratio is 89. Within the framework of ``biased infall'' 
(e.g., Chiappini \& Matteucci 2001), the Galactic disk is slowly formed 
from inside out, which causes gradients in the abundances across the 
disk. The stellar $^{13}$C ejecta, reaching the interstellar medium 
with a time delay, are less dominant in the young stellar disk of the 
outer Galaxy than in the inner Galaxy and the old stellar body of the 
LMC (see, e.g., Hodge 1989 for the star formation history of the LMC). 
The solar system ratio, referring to a younger more $^{13}$C deficient 
disk, is therefore higher than that measured in the present local 
interstellar medium. Consistent with this idea, $^{12}$C/$^{13}$C 
ratios are particularly low ($\sim$25) in the Galactic center region 
with its old bulge (e.g., G{\"u}sten et al. 1985), while inflowing 
or infalling gas from outside appears to be characterized by higher 
ratios (Riquelme et al. 2010). We note that this scenario also 
explains other isotope ratios based on differences of primary and 
secondary nucleosynthesis, like that of $^{16}$O (a product of massive
stars, $\ga$8\,M$_{\odot}$) and $^{17}$O (a product of lower mass stars), 
while $^{18}$O is apparently most efficiently synthesized in metal rich 
stars of large mass (e.g., Wouterloot et al. 2008).

With respect to isotope ratios the extragalactic space beyond the Magellanic 
Clouds is almost unexplored and therefore very interesting to investigate 
(for previous pioneering efforts, see Aalto et al. 1991; Casoli et 
al. 1992; Henkel et al. 1993; Henkel \& Mauersberger 1993; Wang et al. 
2004; Muller et al. 2006; Henkel et al. 2010; Mart\'{\i}n et al. 
2010; Gonz{\'a}lez-Alfonso et al. 2012; Danielson et al. 2013). What ratios 
can be found when observing objects outside the Local Group of galaxies at 
low and high redshifts and in environments, which drastically differ from 
those in the Milky Way and the LMC? Is the Galaxy typical for its class 
or are its isotopic properties exceptional? And what kind of isotopic 
compositions can be expected in optical lines, when trying to determine 
high precision redshifts and to constrain variations in physical constants 
through time and space (e.g., Levshakov et al. 2006)? 

In the following we present and analyze new CN and CO data from the nearby 
prototypical starburst galaxy NGC~253 and the ultraluminous merger Mrk~231, 
in order to derive and to compare the carbon isotope ratios in these different 
environments. Sect.\,2 describes observations and data reduction. Sect.\,3
presents the CN and CO measurements and data analysis toward NGC253, including
carbon and oxygen isotope ratios and CN excitation temperatures. In Sect.\,4 
we discuss our data from Mrk~231 and provide a general overview over 
extragalactic carbon isotope determinations in targets beyond the Magellanic 
Clouds. Sect.\,5 summarizes the main results.

\section{Observations}

The $\lambda$$\sim$3 and 1.3\,mm measurements toward NGC~253 were obtained with 
the IRAM 30-m telescope (project 078--12) at Pico Veleta, Spain\footnote{Based
on observations carried out with the IRAM 30-m telescope. IRAM is supported
by INSU/CNRS (France), the MPG (Germany), and IGN (Spain)} during August 
5 and 6, 2012. Full width to half power beam widths (FWHPs) were about 22$''$ 
and 11$''$. The EMIR SiS receivers were employed with system temperatures of 
$T_{\rm sys}$ $\sim$ 130, 240, 240, and 260~K at 109.6, 113.4 ($\lambda$$\sim$3\,mm), 
221.3, and 225.0\,GHz ($\lambda$$\sim$1.3\,mm) on an antenna temperature scale. Adopted
beam and forward hemisphere efficiencies are 0.80 and 0.95 at $\lambda$ $\sim$ 
3\,mm and 0.62 and 0.92 at $\lambda$ $\sim$1.3\,mm. As backend we used 
Fast Fourier Transform spectrometers with a channel spacing of 195.3\,kHz, 
covering two contiguous 4\,GHz segments in dual linear polarization at 
$\lambda$ $\sim$ 3 and also at $\lambda$ $\sim$ 1.3\,mm. Channel spacings 
are 0.53 and 0.26\,km\,s$^{-1}$, respectively. The spectra were obtained 
with a wobbling secondary mirror using a switch cycle of a few seconds 
(2\,s on-source, 2\,s off-source) and a beam throw of $\pm$100$''$. No 
absorption features are seen in any spectrum, potentially caused by using too 
small a beam throw. The pointing accuracy, based on nearby continuum sources, 
was accurate to $\sim$5$''$. For calibration, see Sect.\,3.6.2. Table~\ref{tab1} 
displays some essential parameters of the observations.

\begin{table}
\caption[]{Observational parameters$^{\rm a)}$}
\begin{flushleft}
\begin{tabular}{ccccccc}
\hline
  Band     & $\nu$    &  $T_{\rm sys}$  & $\theta_{\rm b}$ & $f_{\rm mb}$  & $f_{\rm fh}$  & $S$/$T_{\rm mb}$ \\
  (mm)     & (GHz)    &        (K)      &    ($''$)        &               &               & (Jy/K) \\
\hline 
           &          &                 &                  &               &                    \\
 3         & 109.635  &      130        &     23           &      0.80     &   0.95      &  5.2 \\
 3         & 113.365  &      240        &     22           &      0.80     &   0.95      &  5.1 \\
 1.3       & 221.315  &      240        &     11           &      0.62     &   0.92      &  4.8 \\
 1.3       & 225.045  &      260        &     11           &      0.62     &   0.92      &  4.8 \\
\hline
\end{tabular}
\end{flushleft}
a) Receiver band ($\lambda$ $\sim$ 3 or 1.3\,mm), frequencies ($\nu$), system 
temperatures ($T_{\rm sys}$) on an antenna temperature scale ($T_{\rm A}^*$), full 
width to half power (FWHP) beam widths in units of arcseconds ($\theta_{\rm b}$), 
and adopted main beam efficiencies ($f_{\rm mb}$) and forward hemisphere efficiencies 
($f_{\rm fh}$) for the four 4\,GHz wide frequency intervals (2$\times$8\,GHz), 
simultaneously observed by the 30-m telescope. The last column provides conversion 
factors from main beam brightness temperatures (in Kelvin) to flux density (Jansky) units.
\label{tab1}
\end{table}

Complementary $\lambda$ $\sim$ 3 and 1.3\,mm observations of Mrk~231 were taken 
in January and May 2011 (project 233--11) with the IRAM 30-m telescope using 
the same receivers and observing mode with the WILMA backend under varying 
weather conditions. This backend provided channel spacings of 2\,MHz. 

Data analysis was performed with the GILDAS data reduction 
package\footnote{Grenoble Image and Line Data Analysis Software:
http://www.ira.inaf.it/~brand/gag.html}, revealing excellent baselines
only requiring the subtraction of baselines of order $\leq$2 for both galaxies.
The $\lambda$ = 3 and 1.3\,mm data were taken simultaneously and the 
pointing difference between the two EMIR receivers was found to be $\la$2$''$.

\begin{figure}[h]
\vspace{0.0cm}
\centering
\resizebox{23.0cm}{!}{\rotatebox[origin=br]{-90}{\includegraphics{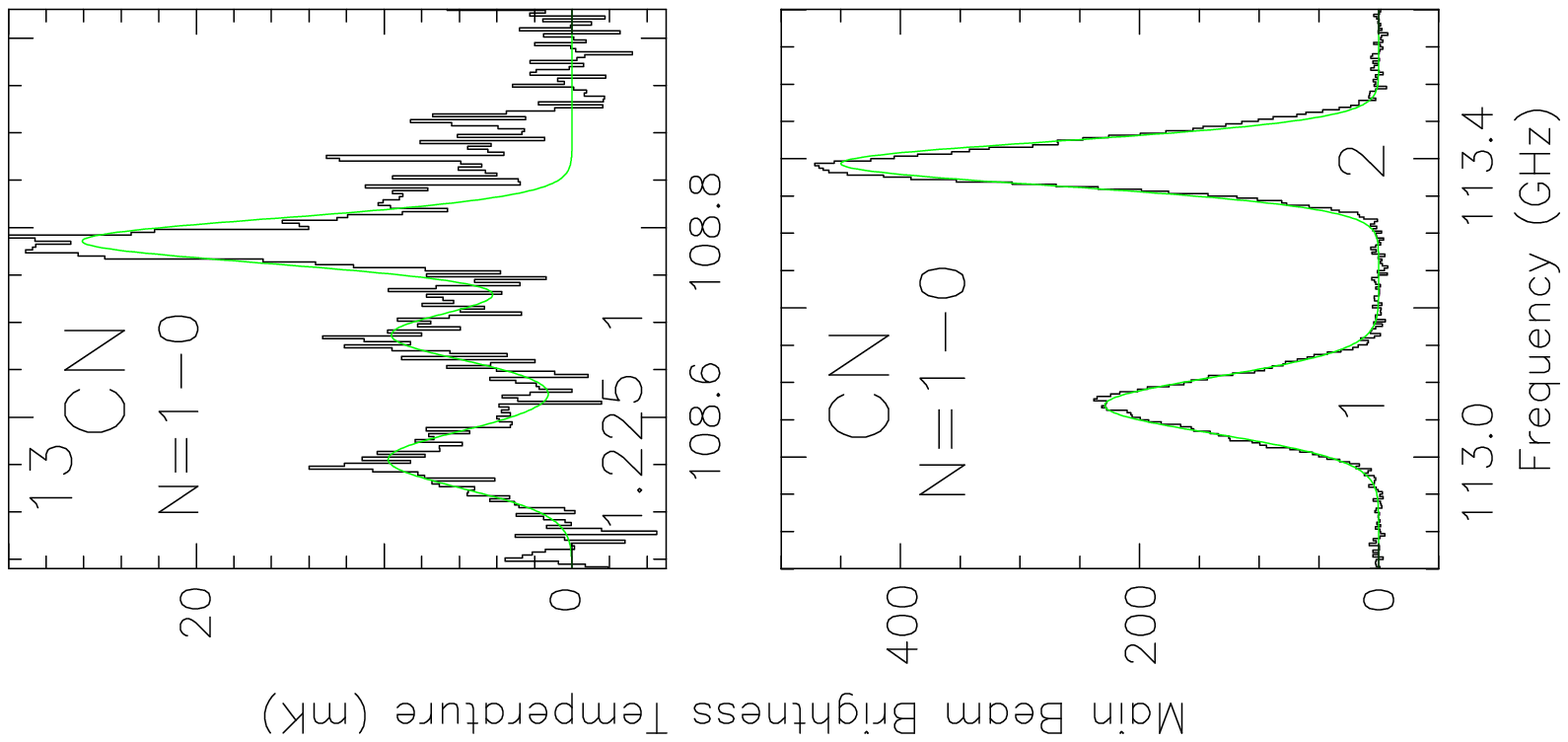}}}
\vspace{-1.0cm}
\caption{CN $N$=1$\rightarrow$0 spectra (black) and Gaussian fits (green) from NGC~253 on a Local Standard 
of Rest (LSR) $V_{\rm LSR}$ = 0\,km\,s$^{-1}$ frequency scale. Both spectra are smoothed to a channel 
spacing of $\sim$8.5\,km\,s$^{-1}$ (3.125\,MHz). {\it Lower panel}: The $J$ = 1/2$\rightarrow$1/2
(left) and $J$ = 3/2$\rightarrow$1/2 (right) groups of CN lines. {\it Upper panel}: The strongest
feature is the 0$_0$$\rightarrow$1$_{-1}$ E line of CH$_3$OH (methanol). Far left: 
the $F_1$=0, $F_2$=1$\rightarrow$0 and $F_1$=1, $F_2$=1$\rightarrow$1 group of $^{13}$CN $N$ 
= 1$\rightarrow$0 lines. In between this spectral feature and the methanol line: 
the $F_1$=1, $F_2$=2$\rightarrow$1 group of $^{13}$CN transitions (for CN and $^{13}$CN rest 
frequencies, see Skatrud et al. 1983 and Bogey et al. 1984). The emission on the right
hand side of the methanol line near 108.9\,GHz might be caused by SiS 6$\rightarrow$5. 
Numbers at the foot of each spectral CN or $^{13}$CN feature provide expected relative 
intensities with respect to the weaker group of lines in case of optically thin emission 
under conditions of local thermodynamical equilibrium. For less sensitive CN spectra 
obtained with smaller bandwidths, see Fig.~1 of Henkel et al. (1993).}
\label{fig1}
\end{figure}

\section{CN and CO toward NGC~253}

\subsection{The galaxy NGC~253}

The Sculptor galaxy NGC~253, an almost edge-on barred spiral ($i$=72$^{\circ}$--78$^{\circ}$;
Pence 1981; Puche et al. 1991), is one of the most prolific infrared and molecular lighthouses
of the entire extragalactic sky. At a distance of $D$ $\sim$ 3\,Mpc (e.g., Mouhcine et al. 2005;
Rekola et al. 2005), it is a prime example of a galaxy with a nuclear starburst devoid of 
an active galactic nucleus (e.g., Ulvestad \& Antonucci 1997; Henkel et al. 2004). Because of the 
exceptional strength of its molecular lines, NGC~253 was selected as the target of choice for the 
first unbiased molecular line survey of an extragalactic source (Mart\'{\i}n et al. 2006). It is 
therefore a highly suitable target for this study.

\begin{table*}
\caption[]{CN line parameters of NGC~253, obtained from Gaussian fits$^{\rm a)}$}
\begin{flushleft}
\begin{tabular}{cccccc}
\hline
Line                                 &$\int T_{\rm mb}$&    $\nu$           &$\Delta\nu_{\rm 1/2}$&    $T_{\rm mb}$\\
                                     &    (K MHz)      &    (MHz)           &  (MHz)     &     (mK)       \\
\hline 
                                     &                 &                    &            &        &                \\
CN $N = 1\rightarrow$0               &21.637$\pm$0.083 &113069.244$\pm$0.166& 88.721$\pm$0.388    & 244$\pm$1      \\
				     &33.537$\pm$0.061 &113393.540$\pm$0.063& 70.041$\pm$0.170    & 479$\pm$1      \\
                                     &                 &                    &                     &                \\
$^{13}$CN $N = 1\rightarrow$0        & 0.736$\pm$0.086 &108555.278$\pm$3.931& 70.565$\pm$9.396    &10.4$\pm$1.8    \\
                                     & 0.632$\pm$0.091 &108686.718$\pm$4.555& 61.822$\pm$10.458   &10.2$\pm$2.3    \\
CH$_3$OH 0$_0$$\rightarrow$1$_{-1}$ E& 1.569$\pm$0.108 &108785.756$\pm$1.333& 56.561$\pm$6.061    &27.7$\pm$3.5    \\
                                     &                 &                    &                     &                \\
CN $N = 2\rightarrow$1               &15.009$\pm$0.160 &226138.445$\pm$0.556&116.099$\pm$1.369    & 129$\pm$2      \\
                                     &29.210$\pm$0.167 &226518.547$\pm$0.324&115.337$\pm$0.750    & 253$\pm$2      \\
                                     &47.099$\pm$0.164 &226676.359$\pm$0.174&105.801$\pm$0.463    & 445$\pm$2      \\
                                     &                 &                    &            &                     &                \\
\hline
\end{tabular}
\end{flushleft}
a) Because of large bandwidths (4--8\,GHz), yielding potentially complex velocity-frequency correlations, all fits were obtained
on a frequency scale. All fitted CN components refer to groups of individual lines. Col.\,3 displays observed frequencies, 
referring to the Local Standard of Rest (LSR) $V_{\rm LSR}$ = 0\,km\,s$^{-1}$ frequency scale. All given errors are standard deviations 
obtained from Gaussian fits. The $T_{\rm mb}$ values (last column) were obtained from the values given in Cols.\,2 and 4. 
Calibration uncertainties are not considered here but are discussed in Sect\,3.6.2. 
\label{tab2}
\end{table*}

\subsection{Our data}

CN spectra are complex. Each CN rotational energy level with $N$$>$0 is split into a doublet 
by spin-rotation interaction. Because of the spin of the nitrogen nucleus ($I_1$=1), each of 
these components is further split into a triplet of states. The $^{13}$CN spectrum is further 
complicated by the spin of the $^{13}$C ($I_2$=1/2) nucleus. 

Figure~\ref{fig1} shows our $\lambda$$\sim$3\,mm CN spectra. The lower panel displays the $N$ = 
1$\rightarrow$0 $J$=1/2$\rightarrow$1/2 (left) and $J$=3/2$\rightarrow$1/2 (right) groups of lines 
of $^{12}$C$^{14}$N (hereafter CN). The upper panel visualizes the blended $N$ = 1$\rightarrow$ 0 $F_2$ = 
1$\rightarrow$0 and 1$\rightarrow$1 (far left) and the slightly weaker $F_2$ = 2$\rightarrow$1 
(next feature to the right) transitions of $^{13}$C$^{14}$N (hereafter $^{13}$CN). This represents 
one of the first detections of $^{13}$CN in extragalactic space (cf. Aladro et al. 2013; for the 
first detection in the local interstellar medium, see Gerin et al. 1984). The $^{13}$CN spectrum 
has an rms noise level of 2\,mK (channel width: 8.5\,km\,s$^{-1}$). While the upper panel of 
Fig.~\ref{fig1} only shows a small spectral segment, we note that the entire spectrum has a width 
of 4\,GHz. Therefore the (flat) baseline and noise level are well defined. Table~\ref{tab2} 
provides the corresponding line parameters. Again it should be emphasized that the Gaussian fit 
result for $^{13}$CN, fitting the two $^{13}$CN profiles and the dominant CH$_3$OH feature 
simultaneously, is very robust. 

A comparison with the approximate rest frequencies of the different 
groups of lines shows that we mainly see the high velocity component of NGC~253 with a recessional 
velocity of $V_{\rm LSR}$ $\sim$ +290\,km\,s$^{-1}$ (e.g., Mart\'{\i}n et al. 2006), located  
several arcseconds south-west of the kinematical center with an extent of order 10$''$
(Mauersberger et al. 1996; Peng et al. 1996; Garc\'{\i}a-Burillo et al. 2000; Paglione et al. 
2004; G{\"u}sten et al. 2006; Lebr{\'o}n et al. 2011; Sakamoto et al. 2011; Bolatto et al. 2013). 
The similarly extended lower velocity component near 170\,km\,s$^{-1}$, mainly arising 
from several arcseconds north-east of the dynamical center, is too weak to be detected at 
significant levels (but see Sect.\,3.8). For the dominant 0$_0$ $\rightarrow$ 1$_{-1}$ E 
line of methanol (CH$_3$OH), we obtain $V_{\rm LSR}$ = (294$\pm$4)\,km\,s$^{-1}$. While this 
inhibits a comparison of the two major molecular lobes near the center of NGC~253, the dominance
of the high velocity component in the spectra is nevertheless positive. It reduces considerably 
the line widths and thus blending of nearby spectral features.

A search for vibrationally excited CN turned out to be unsuccessful. The higher frequency fine structure
components of the $v$ = 1 $N$ = 1$\rightarrow$0 and 2$\rightarrow$1 CN transitions are blended by 
$^{12}$C$^{17}$O $J$ = 1$\rightarrow$0 and 2$\rightarrow$1. For the lower frequency fine structure
components, we obtain with a channel width of 8.5\,km\,s$^{-1}$ $\lambda$ $\sim$ 3 and 1.3\,mm 
1$\sigma$ noise levels of 3 and 4\,mK (15 and 20\,mJy).

\subsection{On the importance of CN}

Toward NGC~253, the $^{12}$C/$^{13}$C carbon isotope ratio has been previously estimated
from CS (Henkel et al. 1993) and C$_2$H (Mart{\'i}n et al. 2010). While the former authors
propose $^{12}$C/$^{13}$C $\sim$40, the latter find $^{12}$C/$^{13}$C $>$ 81. In view of the 
scarcity of $^{12}$C/$^{13}$C determinations from extragalactic sources, this discrepancy 
is is an important issue to resolve. This is one of the main goals of this paper.

As explained in Sect.\,3.2, mm-wave CN spectra contain a multitude of individual 
features. Therefore, a comparison between the tracer species with highest intensities, CO, HCN, 
HCO$^+$, HNC, and CN, clearly favors CN, when attempting to determine optical depths from 
relative line intensities to derive carbon isotope ratios (Henkel et al. 1998). This also 
holds, when including C$_2$H (Mart\'{\i}n et al. 2010), because CN shows the broadest frequency 
coverage of spectral fine structure within its mm-wave transitions, which is sufficient 
even in the case of rotationally broadened lines from an edge-on spiral galaxy (e.g., 
Fig.~\ref{fig1}).

\subsection{Problems related to previous $^{12}$C/$^{13}$C determinations}

The $^{12}$C/$^{13}$C $\sim$ 40 estimate from CS by Henkel et al. (1993) for NGC~253 was based 
on the assumption of a $^{32}$S/$^{34}$S ratio of 23 as measured in the solar system and the 
local interstellar medium (Penzias 1980; Wannier 1980). More recent data, however, indicate 
a strong positive $^{32}$S/$^{34}$S gradient in the Galactic disk (Chin et al. 1996) with 
$^{32}$S/$^{34}$S of order 13.5 in the inner disk and a possibly similar ratio in the nuclear 
starburst environment of another active nearby spiral galaxy, NGC4945 (Wang et al. 2004). 

For a ratio of $^{32}$S/$^{34}$S = 13.5, following the procedure outlined by Henkel et al. (1993), 
we would obtain $^{12}$C/$^{13}$C $\sim$ 23 for NGC~253. More recently, Mart\'{\i}n et al. (2005, 
2006) obtained $I$($^{12}$C$^{32}$S 3--2)/$I$($^{12}$C$^{34}$S 3--2) $\sim$ 5.7 and 
$I$($^{12}$C$^{32}$S 3--2)/$I$($^{13}$C$^{32}$S 3--2) $\sim$ 27 (the ratios were calculated
from their Tables~1, giving integrated intensities, while Mart\'{\i}n et al. (2005) discuss 
$I$($^{12}$C$^{32}$S 3--2)/$I$($^{13}$C$^{32}$S 3--2) = 21$\pm$3 in their Sect.\,3.1, derived
from peak flux densities). With the data from their Tables~1 and $^{32}$S/$^{34}$S = 13.5, we 
obtain, following the same procedure (Henkel et al. 1993), from CS a carbon isotope ratio 
of $^{12}$C/$^{13}$C $\sim$ 70 for NGC~253.

In this context it has to be mentioned that $^{32}$S/$^{34}$S = 8$\pm$2 as suggested by
Mart\'{\i}n et al. (2005) cannot be adopted because it has been derived from the 
$^{12}$C/$^{13}$C $\sim$ 40 ratio proposed by Henkel et al. (1993) under the assumption of 
$^{32}$S/$^{34}$S = 23. However, the sulfur isotope ratio of $>$16, suggested more recently by 
Mart\'{\i}n et al. (2010), is free of such a contradiction. To summarize, the poorly 
constrained sulfur isotope ratio in NGC~253 and the (initially unknown) strong Galactic 
$^{32}$S/$^{34}$S gradient (Chin et al. 1996), allowing for a wide range of ratios at 
least in the Galaxy, inhibit any reliable determination of NGC~253's carbon isotope 
ratio based on CS. In view of the qualitative nature of the above mentioned ratios, a 
systematic analysis of error budgets is not feasible.

\subsection{Estimating $^{12}$CN/$^{13}$CN, excitation temperature, and column density}

\begin{figure}[t]
\vspace{0.0cm}
\centering
\resizebox{13.3cm}{!}{\rotatebox[origin=br]{-90}{\includegraphics{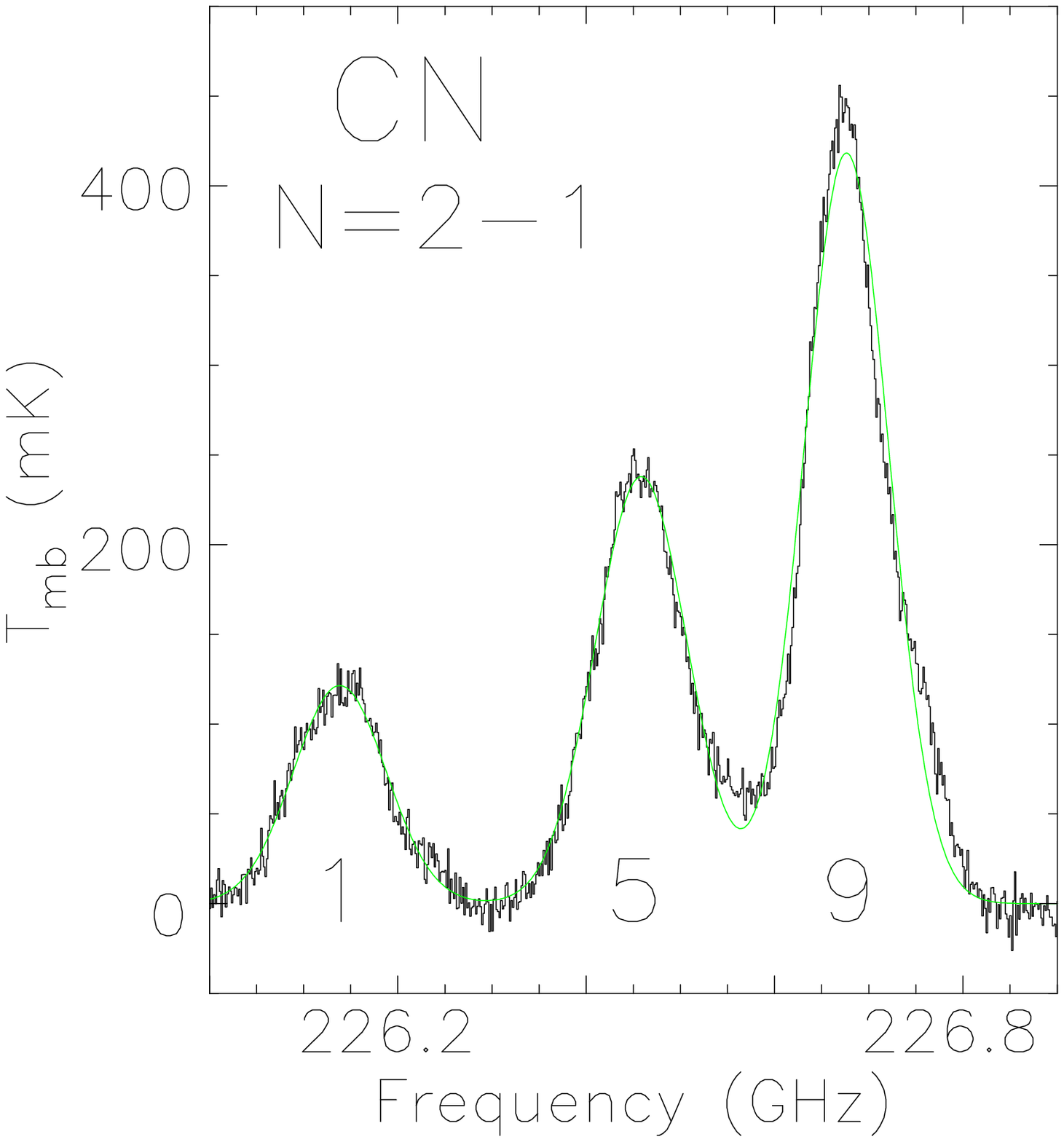}}}
\vspace{-0.4cm}
\caption{The CN $N$=2$\rightarrow$1 spectrum (black) and Gaussian fit (green) of NGC~253 in units 
of main beam brightness temperature on a Local Standard of Rest $V_{\rm LSR}$ = 0\,km\,s$^{-1}$ 
frequency scale. The profile has been smoothed to a channel spacing of $\sim$2.1\,km\,s$^{-1}$ 
(1.5625\,MHz). Numbers at the foot of each spectral feature provide expected relative intensities 
with respect to the weakest group of lines in case of optically thin emission under conditions
of local thermodynamical equilibrium.}
\label{fig2}
\end{figure}

With two detected groups of features in the CN and $^{13}$CN $N$=1$\rightarrow$0 lines 
(Fig.~\ref{fig1}) and three in the CN $N$=2$\rightarrow$1 transition (Fig.~\ref{fig2}), 
we are able to quantify opacity effects without having to rely on a second isotope ratio
(like that of $^{32}$S/$^{34}$S). Therefore, CN provides a good data base to directly estimate 
the $^{12}$C/$^{13}$C ratio in NGC~253. If local thermodynamical equilibrium (LTE) holds 
and lines are optically thin, line intensity ratios should be 1:2 (CN $N$=1$\rightarrow$0), 
1.225:1 ($^{13}$CN $N$=1$\rightarrow$0) and 1:5:9 (CN $N$=2$\rightarrow$1), when moving 
from left to right with increasing frequency in Figs.~\ref{fig1} and \ref{fig2}.

Dividing the integrated intensities of the CN and $^{13}$CN $N$=1$\rightarrow$0 transitions,
we obtain $I$(CN)/$I$($^{13}$CN) = 40. Also accounting for the $F_1$, $F_2$=0$\rightarrow$1 
$^{13}$CN components near 108.4, which are not seen but contribute 7.6\% to the total 
$^{13}$CN emission in the case of optically thin lines and prevailing LTE conditions (see 
Bogey et al. 1984), the ratio drops to $I$(CN)/$I$($^{13}$CN) = 37.5 for the $N$ = 
1$\rightarrow$0 transition. 

To estimate the CN excitation temperature, we note that the $N$ = 2$\rightarrow$1/1$\rightarrow$0
line intensity ratio is 1.655 (Table~\ref{tab2}) on a frequency scale and 0.83 on a velocity scale. 
The latter is relevant here. Accounting for the fact that the CN $N$ = 
2$\rightarrow$1 linear beam size $\theta_{\rm b}$ is half that of the $N$ = 
1$\rightarrow$0  line and assuming that the source size is small with respect to
$\theta_{\rm b, CN 2-1}$, beam dilution is four times higher for the $N$ = 1$\rightarrow$0
than for the 2$\rightarrow$1 transition. The corrected line intensity ratio then becomes
0.21. Fig.~2c of Mauersberger et al. (1996) suggests that the high velocity CO $J$ = 
2$\rightarrow$1 emission has an extent comparable to $\theta_{\rm b, CN 2-1}$. Since
CN may arise from an even more compact region, this provides an upper limit to the possible
extent of the CN emission. Therefore the real CN N = 2$\rightarrow$1/1$\rightarrow$0
intensity ratio may be close to 0.25, but below we will account for the entire range of possible
values. For optically thin emission we use the $N$ = 2$\rightarrow$/1$\rightarrow$0 
ratio to constrain the excitation temperature via 
$$
   0.25 = 4 e^{\rm -x} \times\ 
   \frac{1-e^{\rm -2x}}{1-e^{\rm -x}} \times\ 
   \frac{(e^{\rm 2x}-1)^{-1} - (e^{\rm 2y}-1)^{-1}}{(e^{\rm x}-1)^{-1} - (e^{\rm y}-1)^{-1}}
$$
(e.g., Wang et al. 2004), where x = h$\nu_{10}$/kT$_{\rm ex}$, $\nu_{10}$ = 113.386\,GHz (an
averaged CN $N$ = 1$\rightarrow$0 rest frequency), and y = h$\nu_{10}$/2.73\,k = 1.99. In this 
way we derive an excitation temperature of $T_{\rm ex}$ = 3.5$^{+3.0}_{-0.3}$\,K. The error 
limits account for the entire range of possible line intensity ratios from 0.21 to 0.83. For 
an estimate also discussing effects of line saturation and non-LTE excitation, see Sect.\,3.7.

\subsection{Potential uncertainties in the derived carbon isotope ratio}   

Toward NGC~253, $^{12}$C/$^{13}$C line intensity ratios from carbon bearing species with well detected 
$^{13}$C-containing isotopologues are given by Henkel et al. (1993) for CO $J$ = 1$\rightarrow$0 
and 2$\rightarrow$1, HCN $J$=1$\rightarrow$0, HCO$^+$ $J$=1$\rightarrow$0, and CS 
$J$=3$\rightarrow$2. These ratios do not surpass 20 and can be taken as lower limits in 
view of the unknown but certainly substantial optical depths of the main isotopologues. 
Mart\'{\i}n et al. (2005, 2006) find (as already mentioned) a ratio of 27 for CS
$J$=3$\rightarrow$2. In view of all these data, our determination of the CN/$^{13}$CN
$N$=1$\rightarrow$0 line intensity ratio of order 40 is a major step ahead. 
However, questions related to fractionation, isotope selective photodissociation,
calibration, and optical depths of the main CN species still remain to be discussed.

\subsubsection{Chemical fractionation and isotope selective photodissociation}

Langer et al. (1984) modeled the fractionation of oxygen and carbon in dense 
interstellar clouds with time-dependent chemistry, involving cloud lifetimes up
to 10$^8$\,yr, kinetic temperatures of 6--80\,K, and densities of 5$\times$10$^2$ --
10$^5$\,cm$^{-3}$ for a wide range of metal abundances. While oxygen isotope
fractionation is insignificant under all considered conditions, carbon fractionation
occurs resulting in CO providing too low, HCO$^+$ delivering quite accurate, and CN, 
CS, HCN, and H$_2$CO yielding too high $^{12}$C/$^{13}$C ratios. 

Observations of H$_2$CO, C$^{18}$O, and CN, including the corresponding $^{13}$C-bearing
species, across the Galactic plane (Henkel et al. 1982; Langer \& Penzias 1990; 
Milam et al. 2005) demonstrate how large the numerically predicted discrepancies
may become in the real world. Observationally, the carbon isotope ratios from H$_2$CO
turn out to be larger by $\sim$30\% than those from C$^{18}$O and CN, which are quite
similar. Milam et al. (2005), finding a synthesis for all these data sets, conclude that
there is good agreement between all measured $^{12}$C/$^{13}$C values at a given 
galactocentric radius, independent of the kinetic temperature of the cloud observed. 
Thus chemical fractionation and isotope selective photodissociation do not play a 
dominant role. 

In addition to the advantage of CN showing a peculiar spectral fine structure (see 
Sect.\,3.3), we thus also find that there exists an exemplary Galactic data set 
on CN/$^{13}$CN (Milam et al. 2005). There is no such analog for C$_2$H, encompassing
the Galactic disk and center, which would be relevant for the analysis of NGC~253 by 
Mart\'{\i}n et al. (2010). Since C$_2$H is a molecule, which may also permit the
derivation of line opacities of the main isotopologue in galaxies with moderate line widths, 
a systematic survey of C$_2$H, $^{13}$CCH, and C$^{13}$CH across the Galaxy would be
highly desirable, providing interesting insights into astrochemistry, possibly 
delivering a benchmark for extragalactic data, and allowing us to critically 
compare carbon ratios derived from CN and C$_2$H also in extragalactic sources. In view 
of similar results from H$_2$CO, C$^{18}$O, and CN in the Galaxy (Milam et al. 2005), 
we expect compatible results from C$_2$H as well, but this has still to be demonstrated in a 
rigorous way. Although it would be unexpected (see, e.g., Wang et al. 2009, who find that 
the LMC is well mixed with respect to the carbon isotope ratio), we can also not yet 
firmly exclude that CN and C$_2$H trace different regions with different $^{12}$C/$^{13}$C 
values.

Deviations from relative LTE intensities of different features within the $N$ = 
1$\rightarrow$0 or 2$\rightarrow$1 transitions of C$_2$H have been found to be small (e.g., 
Padovani et al. 2009) suggesting that C$_2$H will be useful for studies of isotope 
ratios. Furthermore, CN and C$_2$H appear to be chemically related. 
Both are common tracers of dark clouds (e.g., Padovani et al. 2009 and references 
therein) and are also probes of photon dominated regions (PDRs; e.g. Simon et al.
1997; Rimmer et al. 2012). Because of high critical densities (for C$_2$H, see Spielfiedel et 
al. 2012; for CN, see Sect.\,3.7), their molecular line emission should be weak in clouds of 
low density. In view of all these common properties, significantly different $^{12}$C/$^{13}$C 
ratios from CN and C$_2$H would be a great surprise.

\subsubsection{Calibration}

Table~\ref{tab1} shows a drastic difference between the system temperatures of the two 
$\lambda$ $\sim$ 3\,mm spectra centered at 109.635 and 113.365\,GHz. This is mainly
a consequence of the extinction caused by atmospheric O$_2$ near 118\,GHz. While 
previous studies were made with comparatively small bandwidths, here we face the 
problem that a single system temperature stands for a spectrum with supposedly 
quite different atmospheric extinctions at its low and high frequency edges. Because
of this, we have to compare our measured main beam brightness temperatures with 
those of previous studies. 

Only data from the 30-m IRAM telescope are considered here. With respect to $^{13}$C$^{16}$O 
(hereafter $^{13}$CO) and $^{12}$C$^{18}$O (hereafter C$^{18}$O) $J$ = 1$\rightarrow$0 
near 110\,GHz (Fig.~\ref{fig3}) agreement with the profiles of Harrison et al. (1999) 
is excellent. When compared with Fig.~1 of Henkel et al. (1993), this also holds for 
the two main groups of features of the CN $N$ = 2$\rightarrow$1 transition near 226\,GHz. 
While our signal-to-noise ratios are much higher, there is also no notable discrepancy 
between Fig.~1 of Henkel et al. (1993) and our $^{13}$CN spectrum near 108.6\,GHz 
(Fig.~\ref{fig1}, upper panel). This mainly refers to the CH$_3$OH 0$_0$$\rightarrow$1$_{-1}$ 
line, because the noise level in the previously published spectrum is too high to detect 
$^{13}$CN. Here we should note that all lines considered so far are well displaced from 
the atmospheric 118\,GHz O$_2$ feature so that atmospheric extinction is not expected 
to vary strongly within the individually observed 4\,GHz wide frequency bands. 

Nevertheless, there are also inconsistencies. Henkel et al. (1988) measured only 
$T_{\rm mb}$ $\sim$ 250\,mK for CN $N$ = 2$\rightarrow$1, which is lower than what 
Henkel et al. (1993) report five years later and what we have observed in this study
($\sim$450\,mK; cf. Table~\ref{tab2}). The IRAM 30-m telescope has been greatly improved during 
the years between 1988 and 1993, so the later observation should be preferred, 
providing support for our spectrum displayed in Fig.~\ref{fig2}. More critical 
is CN $N$ = 1$\rightarrow$0, because it has been observed in the 4\,GHz wide 
band also covering CO $J$ = 1$\rightarrow$0 at the edge of the atmospheric 
118\,GHz O$_2$ feature. Henkel et al. (1988) and Henkel et al. (1993) find 
$T_{\rm mb}$ $\sim$ 300 and 350\,mK for CN $N$ = 1$\rightarrow$0, which is 
$\sim$40\% and 25\% below our value (Table~\ref{tab2}). With the CN $N$ = 1$\rightarrow$0
frequencies being located in between those of $^{13}$CO and C$^{18}$O $J$ = 
1$\rightarrow$0 and CO $J$ = 1$\rightarrow$0, our CO peak intensity is also
relevant for an evaluation of calibration uncertainties. With $T_{\rm mb}$
$\sim$ 5.7 (Fig.~\ref{fig4}), we obtain a value well above that shown in Fig.~5c of 
Mauersberger et al. (1996), reaching only $\sim$4\,K for the $V_{\rm LSR}$ $\sim$
+290\,km\,s$^{-1}$ velocity component seen by us (Sect.\,3.2). This difference of
$\sim$30\% relative to our result is similar to the deviation obtained for CN 
N = 1$\rightarrow$0 and will be implemented as the main uncertainty in our estimate 
of the carbon isotope ratio (see Sect.\,3.7).

\subsubsection{Optical depths}

CN has been observed in several clouds of the Galactic center region (Henkel
et al. 1998), which is as close as we can get to the physical conditions prevailing
in the nuclear region of NGC~253. Toward cloud cores, deviations of the individual 
spectral features from LTE were found to be moderate. $I$(CN)/$I$($^{13}$CN) = 9 -- 
15 in the $N$=1$\rightarrow$0 line, which is below the canonical Galactic center 
ratio of $^{12}$C/$^{13}$C $\sim$ 25, indicating moderate CN saturation. In this
context, we should also mention the interferometric CN observations of the innermost 
4\,pc of our Galaxy (Mart\'{\i}n et al. 2012), which yield carbon isotope ratios of 15--45.
We note, however, that this study addresses a very small region compared to those
discussed here.

In NGC~253, the line intensity ratio of the two $^{13}$CN $N$~= 1$\rightarrow$0
features is 1.165$\pm$0.215, which is consistent with LTE and optically thin emission
(Fig.~\ref{fig1}, upper panel). We may, however, face some moderate saturation in the CN 
$N$~= 1$\rightarrow$0 line. Instead of the 2-to-1 ratio expected under LTE conditions
with optically thin lines, the ratio of integrated intensities between the $J$ = 
3/2$\rightarrow$1/2 and 1/2$\rightarrow$1/2 groups of lines is 1.55$\pm$0.01 (see 
Fig.~\ref{fig1} and Table~\ref{tab2}). We note, however, that the peak line intensity 
ratio between the two groups of line components is almost exactly two. The stronger 
spectral component is narrower, a consequence of the relative frequencies of its 
individual hyperfine components. To get twice the integrated intensity, the stronger 
component would have to have a peak temperature of $T_{\rm mb}$ = 606\,mK, $\sim$2.5 
times the $T_{\rm mb}$ value of the weaker component. With the observed 479\,mK 
instead (Table~\ref{tab2}) and assuming equal excitation temperatures for all 
individual features we can apply the radiative transfer equation to determine the 
optical depth via
$$
        606/479\ \sim\ 1.265\ =\ \frac{\tau}{(1 - e^{-\tau})}.
$$
This gives a peak optical depth of $\tau$ $\sim$ 0.5. For the other measured CN 1$\rightarrow$0 
feature we then obtain an optical depth of $\tau$ $\sim$ 0.2 and effects due to saturation should
amount to $\sim$26.5\% and 10\%, respectively.

\subsection{Consequences}

To determine the carbon isotope ratio, we thus multiply the integrated intensity
$I_{113.1}$ of the weaker CN $N$ = 1$\rightarrow$0 feature by $f_{113.1}$ = 1.1 and 
that of the stronger one, $I_{113.4}$, by $f_{113.4}$ = 1.265 to account for line 
saturation (here and below, the indices refer to redshifted frequencies in units
of GHz). We then obtain with the values given in Table~\ref{tab2} and the correction 
to the measured $I_{\rm 13CN}$ intensity mentioned in Sect.\,3.5 a $^{12}$CN/$^{13}$CN $N$~= 
1$\rightarrow$0 ratio of 
$$
  \frac{(I_{113.1} \times\ f_{113.1}) + (I_{113.4} \times\ 
  f_{113.4})}{1.082 \times\ I_{\rm 13CN}} = 45.
$$
This should be consistent with the carbon isotope ratio (e.g., Milam et al. 2005). 
The main error could be caused by an overestimate of our CN 1$\rightarrow$0
main beam brightness temperature scale by $\sim$30\%, as outlined in Sect. 3.6.2.
This would reduce the isotope ratio to a value of approximately 30. Since all
other errors appear negligible relative to this one, we conclude that {\it the carbon 
isotope ratio is 30--50 in the south-western starburst core of NGC~253}. This is well 
below the lower limit proposed by Mart\'{\i}n et al. (2010), while it is perfectly 
consistent with the ratio derived by Henkel et al. (1993) from CS. Nevertheless, we 
consider this latter agreement as fortuitous. It implies that the $^{32}$S/$^{34}$S sulfur 
isotope ratio in the central part of NGC~253 is close to the local interstellar value, 
in agreement with $>$16, the value suggested by Mart\'{\i}n et al. (2010). While all 
this is highly consistent and straight forward, we still have to emphasize that the 
derived carbon isotope ratio is based on the assumption, that the intrinsic CN $N$ = 
1$\rightarrow$0 relative line strengths are, like those of $^{13}$CN 1$\rightarrow$0, 
close to their LTE values (for further support for this assumption, see Sect.\,4.1).

\begin{figure}[t]
\vspace{0.0cm}
\centering
\resizebox{25.0cm}{!}{\rotatebox[origin=br]{-90}{\includegraphics{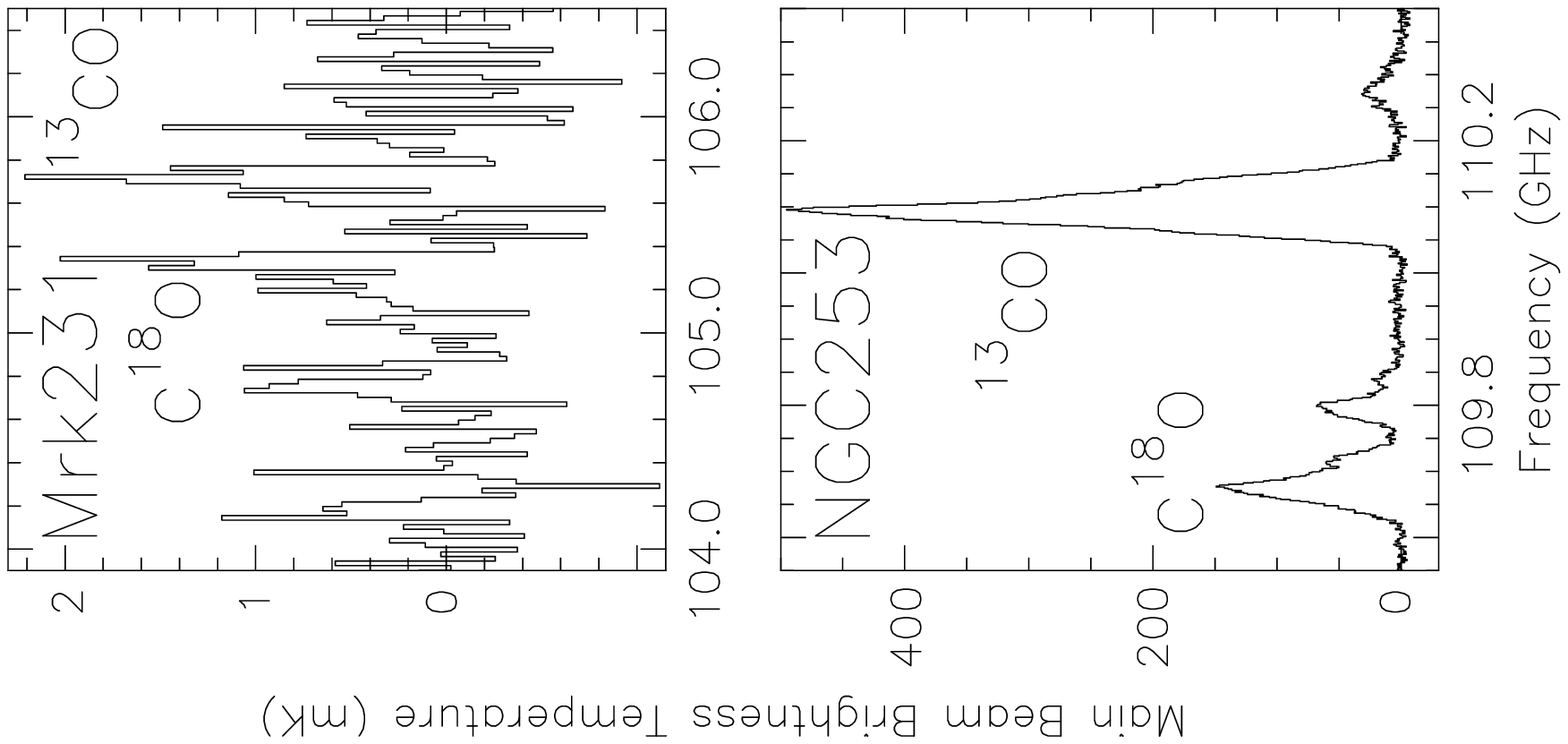}}}
\vspace{-1.0cm}
\caption{C$^{18}$O and $^{13}$CO $J$=1$\rightarrow$0 spectra from NGC~253 (lower panel)
and Mrk~231 (upper panel) on a Local Standard of Rest (LSR) $V_{\rm LSR}$ = 0\,km\,s$^{-1}$ 
frequency scale. The spectra were smoothed to channel spacings of $\sim$4.26 and 
$\sim$60.0\,km\,s$^{-1}$ (1.53 and 21.0\,MHz), respectively. In the lower panel, the 
features at 109.8 and 110.27\,GHz belong to HNCO $J$=5$\rightarrow$4 and CH$_3$CN
$J$=6$\rightarrow$5.}
\label{fig3}
\end{figure}

\begin{figure}[t]
\vspace{0.0cm}
\centering
\resizebox{25.0cm}{!}{\rotatebox[origin=br]{-90}{\includegraphics{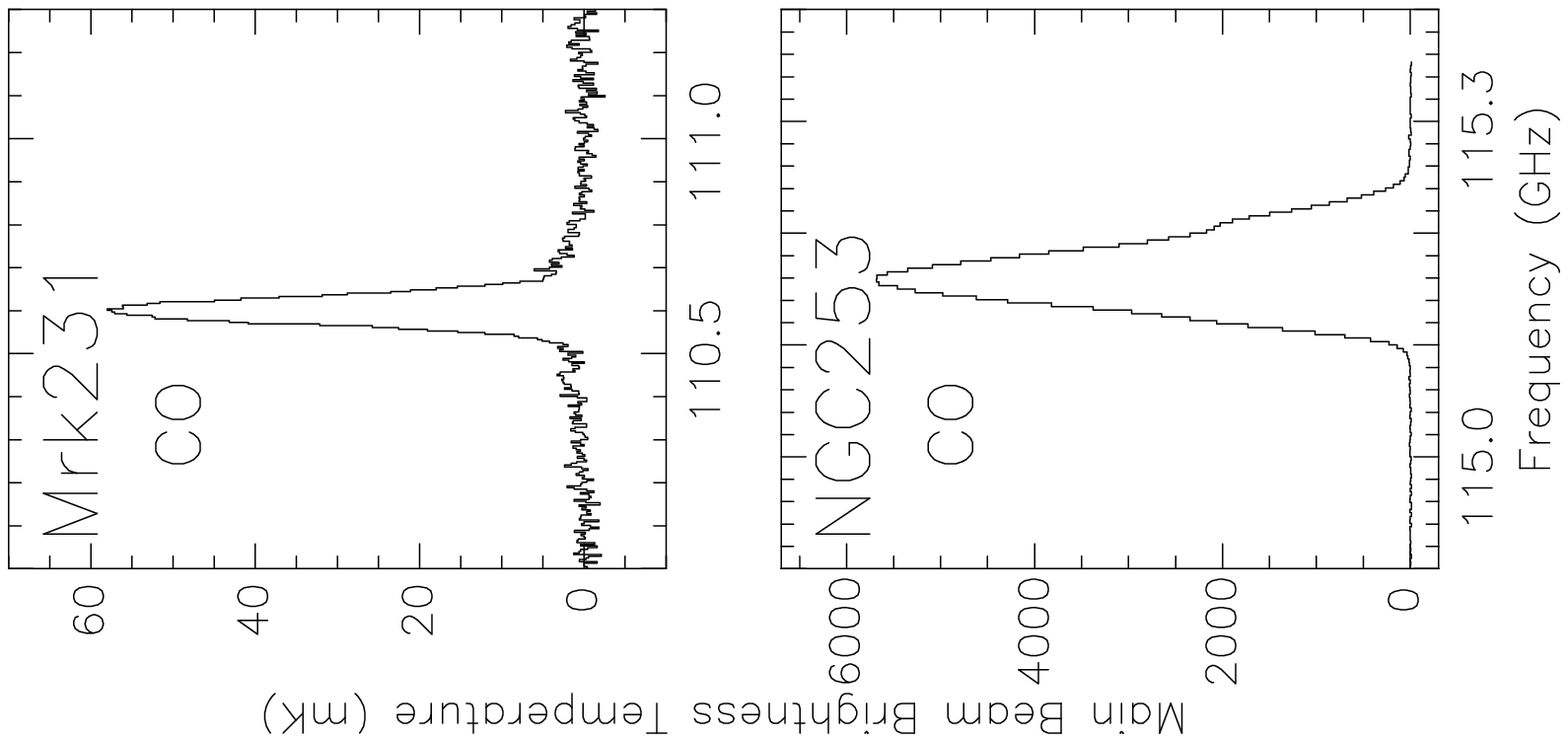}}}
\vspace{-1.0cm}
\caption{CO $J$=1$\rightarrow$0 spectra from NGC~253 (lower panel) and Mrk~231 
(upper panel) on a Local Standard of Rest (LSR) $V_{\rm LSR}$ = 0\,km\,s$^{-1}$ 
frequency scale. The spectra were smoothed to channel spacings of $\sim$8.52 and 
10.84\,km\,s$^{-1}$ (3.125 and 4.0\,MHz), respectively. The frequency range of the
upper panel is wider to show the molecular outflow at the foot of the main spectral
component, seen in CO and other species (e.g., Feruglio et al. 2010; Aalto et al. 2012).}
\label{fig4}
\end{figure}

Accounting with $f_{113.1}$ and $f_{113.4}$ for saturation effects in the CN 1$\rightarrow$0 
transition and neglecting those eventually existing in the $N$ = 2$\rightarrow$1 line, we 
can now also re-evaluate the CN excitation temperature from a modified CN N = 
2$\rightarrow$1/1$\rightarrow$0 intensity ratio, which drops from 0.83 to 0.69 on a velocity 
scale (see Sect.\,3.5). This line ratio would be reduced to 0.172 in case of a point source 
due to the different beam filling factors of the two lines. Following the discussion in Sect.\,3.5, 
and adopting a ratio of 0.2, this results in $T_{\rm ex}$ = 3.2$^{+2.4}_{-0.2}$\,K with the given 
errors covering the entire range of allowed line ratios from 0.172 to 0.69.

\begin{table*}
\caption[]{CO line parameters for NGC~253 and CO and CN line parameters for Mrk~231, obtained from Gaussian fits$^{\rm a)}$}
\begin{flushleft}
\begin{tabular}{crccrr}
\hline
Line            &\multicolumn{1}{c}{$\int T_{\rm mb}$}  &   $V$       &     $\nu$            & $\Delta\nu_{\rm 1/2}$  &     $T_{\rm mb}$    \\
	        &\multicolumn{1}{c}{(K MHz)}            &(km\,s$^{-1}$&     (MHz)            &            (MHz)       &      (mK)           \\
\hline 
                                      &                 &             &                     &                        &                     \\
{\it NGC~253}                         &                 &             &                     &                        &                     \\
$^{12}$C$^{16}$O $J = 1\rightarrow$0  &403.040$\pm$0.125& 285$\pm$1   &115161.571$\pm$0.010 &  68.931$\pm$0.025      & 5847$\pm$3          \\
      Comp. 1                         &360.970$\pm$0.149& 293$\pm$1   &115158.298$\pm$0.002 &  59.460$\pm$0.029      & 6070$\pm$4          \\
      Comp. 2                         & 45.321$\pm$0.121& 155$\pm$1   &115211.774$\pm$0.032 &  35.844$\pm$0.092      & 1264$\pm$5          \\
$^{13}$C$^{16}$O $J = 1\rightarrow$0  & 28.840$\pm$0.070& 285$\pm$1   &110096.675$\pm$0.075 &  63.038$\pm$0.184      &  458$\pm$2          \\
      Comp. 1                         & 24.349$\pm$0.075& 300$\pm$1   &110092.448$\pm$0.072 &  49.949$\pm$0.172      &  487$\pm$2          \\
      Comp. 2                         &  4.593$\pm$0.066& 167$\pm$1   &110139.946$\pm$0.179 &  32.367$\pm$0.358      &  142$\pm$3          \\
$^{12}$C$^{18}$O $J = 1\rightarrow$0  &  8.009$\pm$0.101& 285$\pm$1   &109677.878$\pm$0.372 &  60.921$\pm$0.949      &  131$\pm$3          \\
      Comp. 1                         &  6.652$\pm$0.084& 297$\pm$1   &109673.695$\pm$0.264 &  46.421$\pm$0.691      &  143$\pm$3          \\
      Comp. 2                         &  1.379$\pm$0.073& 169$\pm$2   &109720.494$\pm$0.600 &  29.242$\pm$1.412      &   47$\pm$3          \\
$^{12}$C$^{17}$O $J = 1\rightarrow$0  &  0.629$\pm$0.056& 286$\pm$5   &112252.191$\pm$1.894 &  44.837$\pm$5.007      &   14$\pm$2          \\
                                      &                 &                      &                        &                     \\
$^{13}$C$^{16}$O $J = 2\rightarrow$1  &123.010$\pm$0.261& 278$\pm$1   &220194.289$\pm$0.107 & 105.105$\pm$0.272      & 1170$\pm$4          \\
$^{12}$C$^{18}$O $J = 2\rightarrow$1  & 33.905$\pm$0.212& 278$\pm$1   &219356.938$\pm$0.273 &  92.815$\pm$0.734      &  365$\pm$4          \\
$^{12}$C$^{17}$O $J = 2\rightarrow$1  &  4.216$\pm$0.147& 273$\pm$2   &224509.430$\pm$1.437 &  86.434$\pm$3.724      &   49$\pm$3          \\
				      &                 &             &                     &                        &                     \\
{\it Mrk~231}                         &                 &             &                     &                        &                     \\
$^{12}$C$^{16}$O $J = 1\rightarrow$0  &  4.687$\pm$0.034& 12658$\pm$01&110601.313$\pm$00.268&  75.254$\pm$00.650     & 62.3$\pm$0.7        \\
$^{13}$C$^{16}$O $J = 1\rightarrow$0  &  0.178$\pm$0.041& 12723$\pm$28&105714.712$\pm$09.762&  89.608$\pm$31.488     &  2.0$\pm$0.8        \\
$^{12}$C$^{18}$O $J = 1\rightarrow$0  &  0.142$\pm$0.034& 12685$\pm$31&105325.492$\pm$10.778&  75.964$\pm$26.302     &  1.9$\pm$0.8        \\
				      &                 &             &                     &                        &                     \\
$^{12}$C$^{16}$O $J = 2\rightarrow$1  & 33.178$\pm$0.135& 12661$\pm$01&221196.545$\pm$00.299& 151.903$\pm$00.731     &218.4$\pm$1.4        \\
$^{13}$C$^{16}$O $J = 2\rightarrow$1  &  0.693$\pm$0.142& 12682$\pm$13&211453.470$\pm$09.194&  91.493$\pm$21.645     &  7.6$\pm$2.4        \\
$^{12}$C$^{18}$O $J = 2\rightarrow$1  &  0.544$\pm$0.131& 12691$\pm$16&210643.401$\pm$11.551&  73.979$\pm$22.168     &  7.4$\pm$2.8        \\
				      &                 &             &                     &                        &                     \\
CN $N = 1\rightarrow$0                &  0.295$\pm$0.053& --          &108590.836$\pm$06.522&  84.815$\pm$22.957     &  3.5$\pm$1.1        \\
CN $N = 1\rightarrow$0                &  0.600$\pm$0.040& --          &108898.443$\pm$02.392&  73.780$\pm$05.832     &  8.1$\pm$0.8        \\
				      &                 &             &                     &                        &                     \\
CN $N = 2\rightarrow$1                &  1.714$\pm$0.189& --          &217691.713$\pm$07.215& 139.615$\pm$18.827     & 12.3$\pm$2.1        \\
CN $N = 2\rightarrow$1                &  1.544$\pm$0.197& --          &217442.541$\pm$11.022& 177.146$\pm$25.067     &  8.7$\pm$1.7        \\
CN $N = 2\rightarrow$1                &  0.499$\pm$0.117& --          &217165.557$\pm$08.030&  70.279$\pm$19.209     &  7.1$\pm$2.6        \\
				      &                 &             &                     &                        &                     \\
\hline
\end{tabular}
\end{flushleft}
a) Given frequencies refer to the Local Standard of Rest (LSR) $V_{\rm LSR}$ = 0 km\,s$^{-1}$ frequency 
scale. All errors are standard deviations obtained from Gaussian fits, except those in the last column.
The $T_{\rm mb}$ values and errors were derived by combining the values of columns 2 and 5. Outflow 
components (see Turner et al. 1985; Feruglio et al. 2010; Aalto et al. 2012; Bolatto et al. 2013), 
not apparent in the weaker isotopic lines, are not included in the fits. For potential calibration errors, see 
Sect.\,3.6.2. Because of the large bandwidths of the spectra, yielding potentially complex velocity-frequency 
correlations, frequencies and not velocities are emphasized. For the CO lines, however, also optical (c$z$) 
$V_{\rm LSR}$ values are given, mainly to show the relative importance of the two main spectral features in 
NGC~253. Since each CN component represents a group of hyperfine components of different strength, no velocities 
are given in this case. 
\label{tab3}
\end{table*}

The intensity ratio of the two main CN $N$ = 2$\rightarrow$1 spectral features 
(Fig.~\ref{fig2}) is close to the LTE value of 1.8:1. Their ratio of peak intensities 
is 1.76$\pm$0.02 and the ratio of integrated intensities is 1.61$\pm$0.01, while line 
widths are similar. This suggests that saturation plays only a minor role, affecting 
line intensities by $\la$10\%, as is also suggested by the moderate opacities of the 
1$\rightarrow$0 lines and the low excitation temperature derived above. This effectively 
reduces populations in the higher $N$-levels. While all this is perfectly consistent, the 
intensity of the weakest 2$\rightarrow$1 feature (Fig.~\ref{fig2}) diverges significantly. 
This component appears to be far too strong relative to the others.

An alternative view of the $N$ = 2$\rightarrow$1 line (see Fig.~\ref{fig2}) would be that the 
strongest and weakest $N$ = 2$\rightarrow$1 features are in LTE, while the central feature 
is depleted by a non-LTE effect. Then saturation effects would reduce the main beam brightness 
temperature of the strongest feature from 1274\,mK (nine times the peak intensity of the 
weakest feature multiplied by the ratio of the two line widths, 1.097; see Fig.~\ref{fig2} 
and Table~\ref{tab2}) to the observed 445\,mK, implying a peak optical depth of $\tau$ $\sim$ 3.7. 
Calculating the excitation temperature in this way by adopting for the weaker central feature 
half this peak optical depth, we obtain with the values of Table~\ref{tab2}, i.e. integrating 
over frequency, a $N$ = 2$\rightarrow$1/1$\rightarrow$0 line intensity ratio of 
$$
\frac{(I_{226.7}\times f_{226.7}) + (I_{226.5}\times f_{226.5}) + I_{226.1}}
{(I_{113.1}\times f_{113.1}) + (I_{113.4}\times f_{113.4})}
                          = 3.23.
$$
For the quantities in the denominator, see the previous equation. $f_{226.7}$ = 2.86 and
$f_{226.5}$ = 2.20. Integrating instead over velocity, which is the proper unit, yields 
half the CN line intensity ratio, 1.614, or, for a point source (see Sect.\,3.5), 0.403. The 
resulting excitation then becomes 4.2\,K $<$ $T_{\rm ex}$ $<$ 11.3\,K. Because the CN emission is 
likely not extended with respect to the beam (Sect.\,3.5), the actual excitation temperature 
should be close to 4\,K, still a very low value in spite of all the corrections we have made. 
It is also small in view of the excitation temperatures derived from other species in NGC~253 
(Mart\'{\i}n et al. 2006), strongly indicating subthermal excitation. The critical density 
of CN $N$=1$\rightarrow$0, where collisional excitation rates match those for spontaneous radiative 
decay, is high ($n_{\rm crit}$ $\sim$ 10$^6$\,cm$^{-3}$) and almost as large as that for HCN
$J$=1$\rightarrow$0. Therefore, subthermal emission from a predominantly lower density medium is 
no surprise. The corresponding CN column density is 1.7$\times$10$^{15}$\,cm$^{-2}$ but could be 
up to a factor of 20 higher and almost an order of magnitude lower for the limiting cases $T_{\rm ex}$ 
= 11.3 and 3.0\,K (the latter obtained prior to correct for CN $N$ = 2$\rightarrow$1 line saturation). 
We further note that these column density estimates are speculative, because there may exist 
a hot, dense component with $T_{\rm ex}$ well above 10\,K, which might only become visible when 
observing higher $N$ transitions (for CN chemistry, see, e.g., Simon et al. 1997; Liszt \& 
Lucas 2001). Adopting exclusively collisional excitation and using RADEX (van der Tak et al. 
2007), $T_{\rm ex}$ = 4.0\,K corresponds for kinetic temperatures of 50--100\,K to a density of 
$n$(H$_2$) $\sim$ 2.5 $\times$ 10$^4$\,cm$^{-3}$ This involves Einstein coefficients from Klisch 
et al. (1995), a dipole moment of 1.45\,Debye (Thomson \& Dalby 1968) and He-impact rates from
Lique et al. (2010) scaled by 1.37 to simulate H$_2$.

\begin{figure}[t]
\vspace{-3.8cm}
\centering
\resizebox{25.0cm}{!}{\rotatebox[origin=br]{-90}{\includegraphics{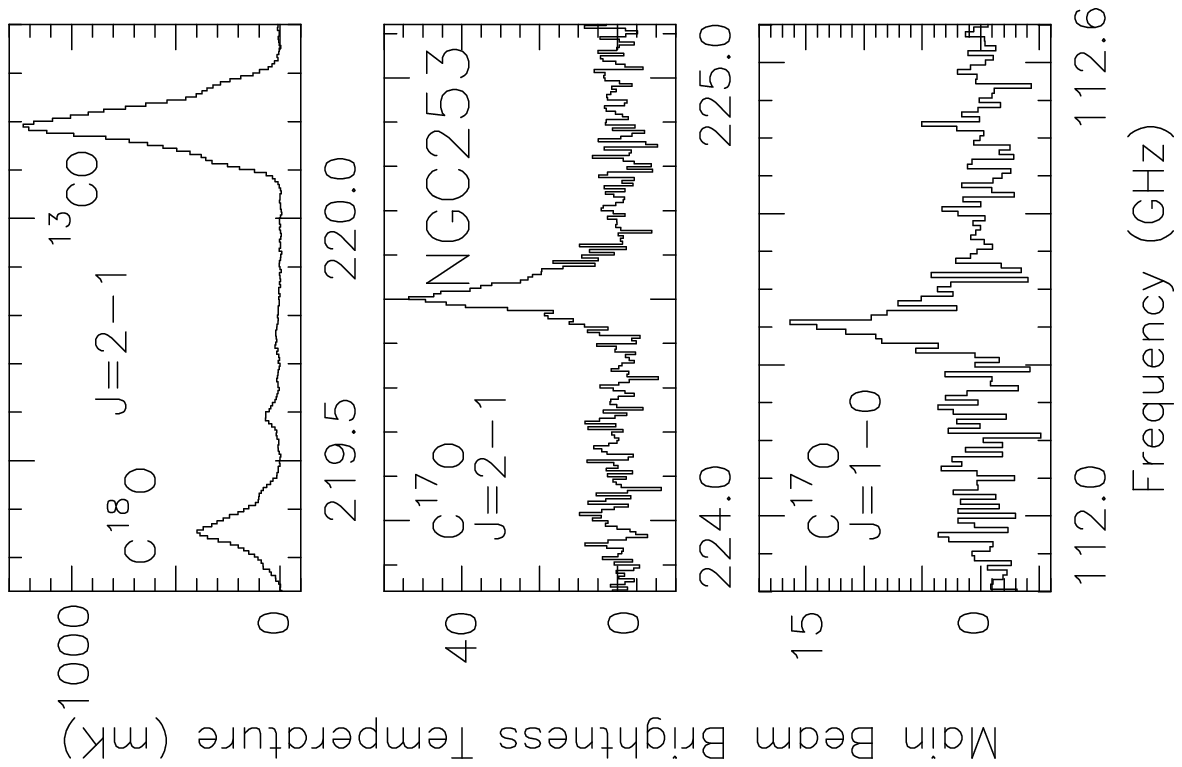}}}
\vspace{-1.0cm}
\caption{$^{13}$CO, C$^{18}$O, and C$^{17}$O spectra from NGC~253 (see also Figs.~\ref{fig3}
and \ref{fig4}) on a Local Standard of Rest (LSR) $V_{\rm LSR}$ = 0\,km\,s$^{-1}$ frequency 
scale. The profiles were smoothed to channel spacings of $\sim$8.5, 8.5, and 17\,km\,s$^{-1}$ 
from top to bottom, which corresponds to 6.25\,MHz.}
\label{fig5}
\end{figure}

\subsection{Beyond CN: Oxygen isotope ratios in NGC~253}

Due to bandwidths of 2$\times$2$\times$4\,GHz (Sect.\,2), our spectra do not only contain CN but
also include a number of CO lines (Figs.~\ref{fig3}--\ref{fig5} and Sect.\,3.6.2). 
Because of their strength and because line shapes are reflecting individual components and 
not groups of hyperfine components (the exception from this rule is $^{12}$C$^{17}$O, 
hereafter C$^{17}$O), not only the $V_{\rm LSR}$ $\sim$ 290\,km\,s$^{-1}$ feature is seen 
in the $J$ = 1$\rightarrow$0 lines (see Sect.\,3.2), but also the lower velocity 
$V_{\rm LSR}$ $\sim$ 170\,km\,s$^{-1}$ component (e.g., Mart\'{\i}n et al. 2006), which 
contributes 10\%-- 20\% to the total line emission in our $\lambda$ = 3\,mm spectra. The 
$J$ = 2$\rightarrow$1 lines obtained with a smaller beam size (Table~1) are best fit by 
a single velocity component. 

Table~3 summarizes the parameters of the Gaussian fits to the line profiles. Noteworthy 
are the extremely high $I$(C$^{18}$O)/$I$(C$^{17}$O) $J$ = 1$\rightarrow$0 and 
2$\rightarrow$1 ratios with respect to the Galactic interstellar medium ($\sim$3.5; 
Wouterloot et al. 2008), 12.7$\pm$1.2 and 8.0$\pm$0.3. These were already noted before 
(e.g., Henkel \& Mauersberger 1993), but the rotational lines of the different isotopologues 
could previously not be observed simultaneously. The high ratios of order 10 were 
interpreted in terms of vigorous massive star formation in a nuclear starburst, containing 
a metal rich gaseous composition.

Adopting $^{12}$C/$^{13}$C = $X$  = 40$\pm$10 from CN (Sect.\,3.7) and keeping in mind that
at least in Galactic star forming regions $^{12}$C/$^{13}$C ratios from CN and C$^{18}$O
are similar (Milam et al. 2005), we can also estimate the optical depths of the various CO 
isotopologues. Only accounting for the errors obtained from Gaussian fits, the CO/$^{13}$CO 
$J$ = 1$\rightarrow$0 line intensity ratio becomes 13.98$\pm$0.34 (Table~3). Since the intensity 
of the CO $J$ = 1$\rightarrow$0 line may be overestimated by 30\% (Sect.\,3.6.2), we estimate 
a ratio of $R$ = 10--14 and obtain with 
$$
        R = \frac{1 - e^{-X\tau}}{1 - e^{-\tau}}
$$
$\tau$(CO 1$\rightarrow$0) = $X$$\tau$ = 1.8 -- 5. This result implies that $^{13}$CO should be 
optically thin, not only in the $J$ = 1$\rightarrow$0 but also in the 2$\rightarrow$1 transition. 
Testing this by comparing the $I$($^{13}$CO)/$I$(C$^{18}$O) line intensity ratios in the two ground 
rotational transitions, we obtain 3.601$\pm$0.046 and 3.628$\pm$0.024, respectively. Within 
the limits of accuracy, the ratios agree with each other, as expected in the case of optically 
thin emission. With the $I$($^{13}$CO)/$I$(C$^{18}$O) value and with 8.9$\pm$1.2 being the weighted mean 
of the $I$(C$^{18}$O)/$I$(C$^{17}$O) ratio from the $J$ = 1$\rightarrow$0 and 2$\rightarrow$1
transitions, $^{16}$O/$^{18}$O = CO/$^{13}$CO $\times$ $^{13}$CO/C$^{18}$O = (40$\pm$10) 
$\times$ (3.62$\pm$0.05) = 145$\pm$36 and $^{16}$O/$^{17}$O = $^{16}$O/$^{18}$O $\times$ 
C$^{18}$O/C$^{17}$O = (145$\pm$36) $\times$ (8.9$\pm$1.2) = 1290$\pm$365 (see also Harrison 
et al. 1999).

\begin{figure}[t]
\vspace{0.0cm}
\centering
\resizebox{25.0cm}{!}{\rotatebox[origin=br]{-90}{\includegraphics{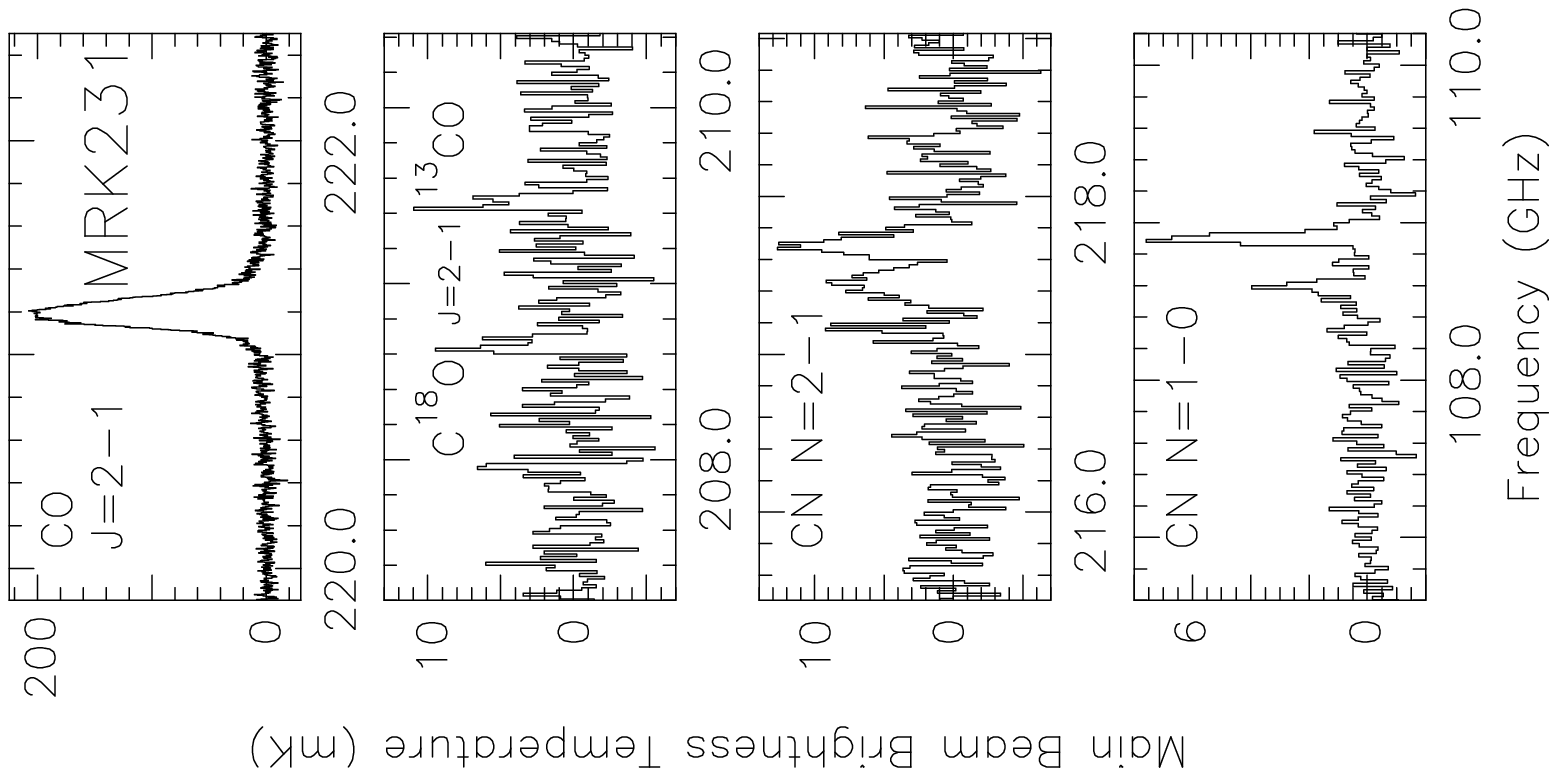}}}
\vspace{-1.0cm}
\caption{CO $J$=2$\rightarrow$1 and CN spectra from Mrk~231 on a Local Standard of 
Rest (LSR) $V_{\rm LSR}$ = 0\,km\,s$^{-1}$ frequency scale. The spectra were smoothed 
to channel spacings of $\sim$2.7, 22, 23, and 60\,km\,s$^{-1}$ ($\sim$2, 16, 16, and 
21\,MHz) from top to bottom. }
\label{fig6}
\end{figure}

\section{Other galaxies}

\subsection{CN in ULIRGs}

In view of the extreme usefulness of CN as a tracer of the carbon isotope ratio 
(Sects.\,3.3 and 3.6.1) and to complement our discussion of the relatively weak starburst 
in NGC~253, it is of interest to gain some idea of $N$ = 1$\rightarrow$0 and 2$\rightarrow$1 
CN emission from a truly luminous local ULIRG (UltraLuminous InfraRed Galaxy). With $L_{\rm IR}$
$\sim$ 2.5$\times$ 10$^{12}$\,L$_{\odot}$ at $z$ = 0.0422 (e.g., Aalto et al. 2012), Mrk~231 
is the target of choice. The two lower panels of Fig.~\ref{fig6} show the spectra, 
Table~\ref{tab3} displays line parameters.

With the enormous infrared luminosity of Mrk~231, hinting at a large mass of dense star 
forming molecular gas, one might naively expect that the CN lines should be more saturated 
than in NGC~253. However, this is not entirely the case. The two components of the 
1$\rightarrow$0 transition have relative intensities exactly as expected for optically 
thin lines under LTE conditions. Excluding the unlikely case that this is an unfortunate 
combination of intrinsic non-LTE line strengths and high optical depth, this provides support 
for our assumption (Sect.\,3.7), that ``intrinsic intensities'' (i.e., intensities after 
removing saturation effects) follow LTE conditions also in NGC~253.  

Nevertheless, the CN spectra from Mrk~231 appear to originate in a different environment
than those from NGC~253. While the peak line temperatures of the $N$ = 1$\rightarrow$0
and 2$\rightarrow$1 lines are similar toward NGC~253, the 2$\rightarrow$1 features are 
significantly stronger than the 1$\rightarrow$0 lines in Mrk~231. Furthermore, the integrated 
line intensity ratios of the three 2$\rightarrow$1 components in Mrk~231 unambiguously indicate 
saturation. Instead of ratios of 1:5:9, expected in the optically thin case (Fig.~\ref{fig2}), 
the integrated intensity ratios are roughly 1:3:3.5 (Fig.~\ref{fig6} and Table~\ref{tab3}). 
Therefore, it is clear that the CN excitation temperature must be higher in Mrk~231 than in 
NGC~253.

For a quantitative estimate, we assume again that the intrinsic intensities (i.e., intensities
in the absence of line saturation) follow LTE conditions (there is no indication for non-LTE 
effects, because features expected to be stronger are stronger and features expected to be 
weaker are indeed weaker; see Fig.~\ref{fig6}). Using the same approach as in Sect.\,3.7, 
this yields opacities of $\tau$ = 0.3$\pm$0.2, 1.4$^{+1.2}_{-0.7}$ and 2.9$^{+1.5}_{-0.9}$ 
for the three CN $N$ = 2$\rightarrow$1 components. To correct relative column densities for 
these optical depths by $\tau$/(1--e$^{-\tau}$), we obtain a factor of 2.3$^{+1.0}_{-0.6}$, 
yielding a modified intensity ratio of $I'$(CN 2$\rightarrow$1)/$I$(CN 1$\rightarrow$0) = 
9.73$^{+4.35}_{-2.31}$. Since we integrated over frequency and not yet velocity, this value 
has to be reduced by a factor of two. Accounting for the different beam sizes at $\lambda$ 
= 3 and 1.3\,mm and realizing that Mrk~231 is spatially unresolved in all of the 30-m beams 
(e.g., Aalto et al. 2012), the final ratio is $I$(CN 2$\rightarrow$1)/$I$(CN 1$\rightarrow$0) 
= 1.22$^{+0.54}_{-0.29}$. Adopting the procedure outlined in Sect.\,3.5, this yields an excitation 
temperature of $T_{\rm ex}$ = 8.4$^{+3.8}_{-1.6}$\,K and a tentative 22$''$ beam averaged column 
density of $N$(CN) $\sim$ 3.4$^{+7.4}_{-1.7}$ $\times$ 10$^{14}$\,cm$^{-2}$. Using RADEX (see
Sect.\,3.7), the corresponding density becomes $n$(H$_2$) $\sim$ 8$\times$10$^4$\,cm$^{-3}$. 
While uncertainties are large, this hints at an excitation temperature about twice as large 
as that in NGC~253 (Sect.\,3.5) and NGC~4945 (Wang et al. 2004) and indicates that CN 
transitions with quantum numbers $N$ $>$ 2 may be of interest, at least in ULIRGs.

\subsection{The carbon isotope ratio in starbursts across the universe}

As we have seen (Sect.\,3.7), the $^{12}$C/$^{13}$C ratio of the nuclear starburst in NGC~253 is, at 
least for the high velocity ($V_{\rm LSR}$ $\sim$ 290\,km\,s$^{-1}$; Sect.\,3.2) component, 
higher than in our Galactic center region. The presence of a bar (e.g., Engelbracht et al. 1998), 
likely providing significant inflow, may be a crucial factor that directs large quantities of fresh 
and poorly proccessed gas toward the nuclear region of NGC~253. While such gas is also reaching 
the central part of our Galaxy (Riquelme et al. 2010), this process may occur on a much larger 
scale in NGC~253, where even indications of massive molecular feed back have already been detected
(Turner 1985; Bolatto et al. 2013). For ongoing star formation, such inflowing gas with high 
$^{12}$C/$^{13}$C ratios may then become even more enriched in $^{12}$C by the 
material ejected from young massive stars. 

Are all starburst galaxies alike with respect to their carbon isotope ratio? Here 
we may differentiate between starbursts in their early and late stages of evolution 
as well as between weak and strong starbursts (see, e.g., Fig.~1 in Mao et al. 2010), 
the latter leading to the presence of (ultra)luminous infrared galaxies ((U)LIRGs). 
Finally, we may also distinguish between galaxies in the local and in the early universe. 

NGC~253 has been believed to host a young starburst (e.g., Garc\'{\i}a-Burillo et al. 
2000; Wang et al. 2004), but in view of detected large-scale outflows (Turner
1985; Bolatto et al. 2013), an intermediate stage of evolution appears to be more likely.  
With a total infrared luminosity of $L_{\rm IR}$ $\sim$ 3$\times$10$^{10}$\,L$_{\odot}$ (e.g., 
Henkel et al. 1986), its level of activity is at the low end of the range observed in starbursts 
of spiral galaxies. A comparison of the carbon isotope ratio in NGC~253 with a starburst in a 
late stage of evolution of similar infrared luminosity, M~82, is not yet possible. Henkel et al. (1998) 
studied M~82 and proposed an isotope ratio of $^{12}$C/$^{13}$C $>$ 40 based on CN and Mart\'{\i}n 
et al. (2010) reported a value $>$138 from C$_2$H, but both results should be taken with 
some degree of scepticism, because $^{13}$CN and $^{13}$CCH or C$^{13}$CH were not detected. 

Mrk~231, one of the most luminous galaxies within a billion lightyears from Earth, contains
an active galactic nucleus (AGN) and may host a starburst in a late stage of evolution. 
This can be deduced from the presence of only one nucleus in this galaxy merger and intense 
outflows of ionized and molecular gas (e.g., Feruglio et al. 2010; Aalto et al. 2012), 
rapidly exhausting the molecular star forming fuel in the central region. Figs.~\ref{fig3}, 
\ref{fig4}, and \ref{fig6} show CO, $^{13}$CO, and C$^{18}$O profiles. Comparing NGC~253 
with Mrk~231, we note that the CO and C$^{18}$O $J$ = 1$\rightarrow$0 peak temperatures are 
$\sim$100 times higher in NGC~253, while for $^{13}$CO $J$ = 1$\rightarrow$0, the ratio
exceeds 200. $I$($^{12}$CO)/$I$(C$^{18}$O) values are 30--40 in in both 
sources. However, the $^{12}$CO/$^{13}$CO line ratios are quite different, with 
$I$($^{12}$CO)/$I$($^{13}$CO)) = 10--14 in NGC~253 (Sect.\,3.8) and 25--50 in Mrk~231 
(Table~\ref{tab3}). While the $^{12}$C/$^{13}$C ratio in the nuclear region of NGC~253 is 
already higher than in our Galactic center region, it appears to be even higher in Mrk~231. 
In NGC~253, $^{13}$CO is much stronger than C$^{18}$O in both ground rotational lines. In 
Mrk~231 both lines show similar intensities (Figs.~\ref{fig3} and \ref{fig6}) and should be 
optically thin in view of their weakness relative to CO. 

Overall, assuming that the CO/C$^{18}$O abundance ratios are the same, Mrk~231 should
have a deficit in $^{13}$C by a factor of almost three relative to NGC~253 (for a statistical 
evaluation comprising many galaxies, see Taniguchi \& Ohyama 1998; Taniguchi et al. 1999), 
possibly yielding $^{12}$C/$^{13}$C $\sim$100 and thus also $^{16}$O/$^{18}$O $\sim$100. 
Interestingly, Greve et al. (2009) and Mart\'{\i}n et al. (2011) find for the less evolved 
merger Arp~220 also $I$($^{13}$CO) $\sim$ $I$(C$^{18}$O). Furthermore, Gonz{\'a}lez-Alfonso et 
al. (2012) derive $^{16}$O/$^{18}$O $\sim$ 100 from OH Herschel data for the same source. 
Arp~220 still possesses two well separated nuclei and shows not yet any outflow which could 
match that seen in Mrk~231. In view of their different stages of evolution, it is therefore 
surprising that Arp~220 and Mrk~231 can be characterized by quite similar carbon and 
$^{16}$O/$^{18}$O ratios. Even the utraluminous eyelash galaxy at redshift 2.3, the first 
high-$z$ galaxy with detected C$^{18}$O emission, shows similar $^{13}$CO and C$^{18}$O 
intensities (Danielson et al. 2013), indicating a $^{13}$C depletion with respect to local
more quiescent galaxies. The LIRG NGC~1068 with an uncorrected $I$($^{12}$CN)/$I$($^{13}$CN) 
ratio of $\sim$50 (Aladro et al. 2013) might be an intermediate case (however, its $^{13}$CN 
features are weaker and therefore show lower signal-to-noise ratios than those displayed in 
our Fig.~\ref{fig1}). The Cloverleaf quasar at redshift $z$ $\sim$ 2.5 appears to be even more 
extreme. Based on a large number of CO data and the first detection of $^{13}$CO at high redshift, 
the $^{12}$C/$^{13}$C ratio should be well above 100 (Henkel et al. 2010). 

A summary of these results is given in Table~\ref{tab4}, where sources with different properties 
are listed together with their carbon isotope ratio. This is a first attempt to set up such a
table. In the Galaxy, there is not only a carbon isotope ratio gradient, but there are 
also indications for dispersion at a given galactocentric radius (e.g., Milam et al. 2005), 
which is not unexpected in view of radial gas streaming and potential cloud-to-cloud variations 
due to local supernovae or ejecta by late-type stars. With respect to external galaxies,
we are still far away from such a level of precision. More accurate determinations of the 
carbon isotope ratio in the galaxies listed in Table~\ref{tab4} as well as in other 
extragalactic targets, also including other classes of objects, are urgently needed. An obvious 
example would be M~82 as the prototype for a weak starburst at a late stage of evolution. 
Submillimeter galaxies (SMGs) would also be attractive. 

The few determined values (Table~\ref{tab4}) indicate a trend with high-$z$ ULIRGs showing the highest 
carbon isotope ratios. Low-$z$ ULIRGs appear to contain more processed material but still show 
values near $^{12}$C/$^{13}$C $\sim$ 100. The large values relative to those found in most parts
of the Milky Way indicate that (1) the bulk of the material originates from a massive inflow
of poorly processed $^{13}$C deficient gas and/or that (2) there is a large input of $^{12}$C-rich 
gas from ejecta of massive stars. The latter may become more dominant, if the number of such
stars would be enhanced by a top-heavy stellar initial mass function, a result of the high
kinetic temperatures expected in extreme cosmic-ray dominated environments (Papadopoulos et al.
2011). Weaker starbursts may show a moderate enhancement over the classical value for the Galactic 
center region, but all these statements are still based on a rather small amount of sources. Right 
now, we are only beginning to collect the required data for a comprehensive understanding of CNO and 
S/Si nucleosynthesis based on extragalactic molecular spectra.

\begin{table}
\caption[]{Extragalactic carbon isotope ratios$^{\rm a)}$}
\begin{flushleft}
\begin{tabular}{ccrc}
\hline
Class                              &      Target    &  $^{12}$C/$^{13}$C  & Ref. \\
\hline 
                                   &                &                     &      \\
Quiescent spiral, center           & Milky Way      &     $\sim$25        & 1    \\
Low level starburst                & NGC253         &     $\sim$40        & 2    \\
Evolved local ULIRG                & Mrk~231/Arp~220&     $\sim$100       & 2,3  \\
Redshift 2.5 ULIRG                 & Cloverleaf     &     $>$100          & 4    \\
                                   &                &                     &      \\
\hline
\end{tabular}
\end{flushleft}
a) Note that the carbon isotope ratios become less certain from top to bottom.
While the ratio in the central molecular zone of our Galaxy is well established,
the ratio for the Cloverleaf quasar is much less constrained. References (last column): 
[1] G{\"u}sten et al. (1985), [2] this paper, [3] Gonz{\'a}lez-Alfonso et al. 2012,
[4] Henkel et al. 2010.
\label{tab4}
\end{table}

\section{Conclusions}

Using the IRAM 30-m telescope at Pico Veleta, we have detected two CN isotopologues
toward the nearby starburst galaxy NGC~253 and four and three CO isotopologues 
toward NGC~253 and the ultraluminous merger galaxy Mrk~231, respectively. CN
$N$=1$\rightarrow$0 and 2$\rightarrow$1 spectra from Mrk~231 are also 
presented. The main results of this study are:

\begin{itemize}

\item CN appears to be the best tracer to determine carbon isotope 
ratios in nearby external galaxies.

\item Toward NGC~253, the measured $^{13}$CN $N$ = 1$\rightarrow$0 line 
intensities are compatible with local thermodynamical equilbrium (LTE) under 
optically thin conditions. The relative line intensities of the 
$^{12}$CN $N$ = 1$\rightarrow$0 features are best explained by 
LTE conditions modified by moderate saturation, affecting
the peak intensity of the weaker component by $\sim$10\% and the 
stronger component by $\sim$25\%. For $^{12}$CN 2$\rightarrow$1,
either the weakest of the three observed line components is
enhanced or the feature of intermediate intensity is depleted
relative to the expected LTE intensity under optically
thin conditions.

\item Accounting for calibration uncertainties and moderate saturation
in the $^{12}$CN 1$\rightarrow$0 line, the $^{12}$C/$^{13}$C isotope ratio
becomes 40$\pm$10 for the molecular core peaking some arcseconds
south-west of the dynamical center of NGC~253. Combined with data from 
several CO isotopologues and adopting this $^{12}$C/$^{13}$C ratio also 
for the CO emitting gas (which is supported by results from Galactic CN 
and C$^{18}$O data), this yields $^{16}$O/$^{18}$O = 145$\pm$36 and 
$^{16}$O/$^{17}$O = 1290$\pm$365. 

\item CN and C$_2$H show both a number of hyperfine components, which allows 
us to determine optical depths even in extragalactic spectra covering a broad
velocity range. A systematic survey of C$_2$H and its $^{13}$C bearing 
isotopologues in star forming clouds of the Galaxy would thus be essential 
to check whether resulting carbon isotope ratios are consistent with those 
already derived from H$_2$CO, C$^{18}$O, and CN.

\item Toward NGC~253, there is no indication for vibrationally excited CN.
The lower frequency fine structure components in the $v$ = 1 $N$ = 1$\rightarrow$0 
and 2$\rightarrow$1 transitions are not seen down to rms levels of 3 and 4\,mK (15 
and 20\,mJy) in 8.5\,km\,s$^{-1}$ wide channels. Those at higher frequency are 
blended by C$^{17}$O.

\item The CN excitation temperature in NGC~253, derived from the $N$ = 
1$\rightarrow$0 and 2$\rightarrow$1 lines is 3--11\,K, with a most likely 
value of $T_{\rm ex}$ $\sim$ 4\,K. With this value, the column density 
becomes $N$(CN) = 2 $\times$ 10$^{15}$\,cm$^{-2}$ and the density, assuming
purely collisional excitation, becomes $n$(H$_2$) $\sim$ 2.5$\times$10$^4$\,cm$^{-3}$.

\item CN data from the ultraluminous merger Mrk~231 indicate that the 
excitation temperature is enhanced by a factor of two with respect to 
NGC~253 and NGC~4945. In Mrk~231, relative CN line intensities within the 
$N$ = 1$\rightarrow$0 and 2$\rightarrow$1 transitions are compatible with 
local thermodynamical equilibrium. While the 1$\rightarrow$0 transitions
appear to be optically thin, the 2$\rightarrow$1 lines show significant
saturation effects. In view of the excitation temperature, which indicates
a density of almost 10$^{5}$\,cm$^{-3}$ assuming exclusively collisional excitation, 
it would make sense to observe CN transitions with higher quantum numbers $N$ 
in Mrk~231 and other ultraluminous infrared galaxies (ULIRGs).

\item A comparison between NGC~253 and Mrk~231 shows that $^{13}$C$^{16}$O
is underabundant in Mrk~231 relative to $^{12}$C$^{16}$O and $^{12}$C$^{18}$O
by almost a factor of three. This would yield $^{12}$C/$^{13}$C $\sim$ 100 and,
because $^{13}$CO and C$^{18}$O show similar intensities in both the 
$J$ = 1$\rightarrow$0 and 2$\rightarrow$1 lines, also $^{16}$O/$^{18}$O
$\sim$ 100. This is similar to the values determined for Arp~220, even
though Arp~220 is a much less evolved ultraluminous merger.

\item Obtaining a synthesis of so far obtained carbon isotope ratios
from the central regions of actively star forming galaxies, the observed
range of values appears to encompass a full order of magnitude. From
ultraluminous galaxies at high redshift to local ULIRGs, to weaker local
starbursting galaxies and to the central molecular zone of the Milky Way,
ratios are $>$100, $\sim$100, $\sim$40, and 25, respectively. While
this matches qualitative expectations of decreasing $^{12}$C/$^{13}$C
values with time and metallicity, we note that (1) the extragalactic
values are based on an extremely small data base and that (2) the 
ratios for the ULIRGs at high and low $z$ are still rather uncertain.
Furthermore, it still has to be evaluated in how far $^{13}$C-deficient
gas from the outer galactic regions and $^{12}$C-rich ejecta from massive 
stars in a nuclear starburst (the latter possibly enhanced by a top-heavy 
initial mass function), are contributing to raise the carbon isotope ratios 
during the lifetime of a starburst.

\end{itemize}

\acknowledgements
We wish to thank the IRAM staff at the 30-m for their help with the observations
and C.~M. Walmsley and an anonymous referee for carefully reading the manuscript.
Some of the work by CH has been carried out while visiting the ESO-ALMA 
group in Santiago de Chile. ALRD acknowledges an STFC studentship (ST/F007299/1).


\begin{thebibliography}{}
 \bibitem[1991]{xyz}
  Aalto, S., Black, J.~H., Johansson, L.~E.~B., \& Booth, R.~S. 1991, A\&A, 249, 323 
 \bibitem[2012]{xyz}
  Aalto, S., Garc\'{\i}a-Burillo, S., Muller, S., et al. 2012, A\&A, 537, A44 
 \bibitem[2012]{xyz}
  Abia, C., Palmerini, S., Busso, M., \& Cristallo, S. 2012, A\&A, 548, A55
 \bibitem[2013]{xyz}
  Aladro, R., Viti, S., Bayet, E., et al. 2013, A\&A, 549, A39
 \bibitem[1984]{xyz}
  Bogey, M., Demuynck, C., \& Destombes, J.~L. 1984, Can. J. Phys. 62, 1248
 \bibitem[2013]{xyz}
  Bolatto, A.~D., Warren, S.~R., Leroy, A.~K., et al. 2013, Nature, 499, 450
 \bibitem[1992]{xyz}
  Casoli, F., Dupraz, C., \& Combes, F. 1992, A\&A, 264, 55
 \bibitem[2001]{xyz}
  Chiappini, C., Matteucci, F., \& Romano, D. 2001, ApJ, 554, 1044
 \bibitem[1996]{xyz}
  Chin, Y.-N., Henkel, C., Whiteoak, J.~B., Langer, N., \& Churchwell, E.~B. 1996, A\&A, 305, 960 
 \bibitem[2013]{xyz}
  Danielson, A.~L.~R., Swinbank, A.~M., Smail, I. et al. 2013, MNRAS, 436, 2793
 \bibitem[1998]{xyz}
  Engelbracht, C.~W., Rieke, M.~J., Rieke, G.~H., Kelly, D.~M., \& Achtermann, J.~M. 1998, ApJ, 505, 639
 \bibitem[2010]{xyz}
  Feruglio, C., Maiolino, R., Piconcelli, E. et al. 2010, A\&A, 518, L155
 \bibitem[2000]{xyz}
  Garc\'{\i}a-Burillo, S., Mart\'{\i}n-Pintado, J., Fuente, A., \& Neri, R. 2000, A\&A, 355, 499
 \bibitem[1984]{xyz}
  Gerin, M., Combes, F., Encrenaz, P. et al. 1984, A\&A, 136, L17
 \bibitem[2012]{xyz}
  Gonz{\'a}lez-Alfonso, E., Fischer, J., Graci{\'a}-Carpio, et al. 2012, A\&A, 541, A4
 \bibitem[1999]{xyz}
  Greve, T.~R., Papadopoulos, P.~P., Gao, Y., \& Radford, S.~J.~E. 2009, ApJ, 692, 1432
 \bibitem[1985]{xyz}
  G{\"u}sten, R., Henkel, C., \& Batrla, W. 1985, A\&A, 149, 195
 \bibitem[2006]{xyz}
  G{\"u}sten, R., Philipp, S.~D., Wei{\ss}, A., \& Klein, B. 2006, A\&A, 454, L115
 \bibitem[1999]{xyz}
  Harrison, A., Henkel, C., \& Russel, A. 1999, MNRAS, 303, 157
 \bibitem[1993]{xyz}
  Henkel, C., \& Mauersberger, R. 1993, A\&A, 274, 730
 \bibitem[1982]{xyz}
  Henkel, C., Wilson, T.~L., \& Bieging, J. 1982, A\&A, 109, 344
 \bibitem[1985]{xyz}
  Henkel, C., G{\"u}sten, R., \& Gardner, F.~F. 1985, A\&A, 143, 148
 \bibitem[1986]{xyz}
  Henkel, C., Wouterloot, J.~G.~A., \& Bally, J. 1986, A\&A, 155, 193
 \bibitem[1988]{xyz}
  Henkel, C., Mauersberger, R., \& Schilke, P. 1988, A\&A, 201, L23
 \bibitem[1993]{xyz}
  Henkel, C., Mauersberger, R., Wiklind, T., et al. 1993, A\&A, 268, L17
 \bibitem[1994]{xyz}
  Henkel, C., Wilson, T.~L., Langer, N., Chin, Y.-N., \& Mauersberger, R. 1994, 
  in {\it The Stucture and Content of Molecular Clouds}, Lecture Notes in Physics, 
  439, p72
 \bibitem[1998]{xyz}
  Henkel, C., Chin, Y.-N., Mauersberger, R., \& Whiteoak, J.~B. 1998, A\&A, 329, 443
 \bibitem[2004]{xyz}
  Henkel, C., Tarchi, A., Menten, K.~M., \& Peck, A.~B. 2004, A\&A, 414, 117 
 \bibitem[2010]{xyz}
  Henkel, C., Downes, D., Wei{\ss}, A., Riechers, D., \& Walter, F. 2010, A\&A, 516, A111
 \bibitem[1989]{xyz}
  Hodge, P. 1989, ARA\&A, 27, 139
 \bibitem[1995]{xyz}
  Klisch, E., Klaus, T., Belev, S.~P., Winnewisser, G., Herbst, E. 1995, A\&A, 304, L5
 \bibitem[1984]{xyz}
  Langer, W.~D., Graedel, T.~E., Frerking, M.~A., \& Armentrout, P.~B. 1984, ApJ, 277, 581
 \bibitem[1990]{xyz}
  Langer, W.~D., \& Penzias, A.~A. 1990, ApJ, 357, 477
 \bibitem[2011]{xyz}
  Lebr{\'o}n, M., Mangum, J.~G., Mauersberger, R., et al. 2011, A\&A, 534, A56
 \bibitem[2006]{xyz}
  Levshakov, S.~A., Centuri{\'o}n, M., Molaro, P., \& Kostina, M.~V.  2006, A\&A, 447, L21
 \bibitem[2010]{xyz}
  Lique, F., Spielfiedel, A., Feautrier, N. et al. 2010, J. Chem. Phys. 132, 024303
 \bibitem[2001]{xyz}
  Liszt, H. \& Lucas, R. 2001, A\&A, 370, 576
 \bibitem[2010]{xyz}
  Mao, R.-Q., Schulz, A., Henkel, C., et al. 2010, A\&A, 724, 1336
 \bibitem[2006]{xyz}
  Mart\'{\i}n, S., Mart\'{\i}n-Pintado, J., Mauersberger, T., Henkel, C., \& Garc\'{\i}a-Burillo, S. 2005, ApJ, 620, 210
 \bibitem[2006]{xyz}
  Mart\'{\i}n, S., Mauersberger, R., Mart\'{\i}n-Pintado, J., Henkel, C., \& Garc\'{\i}a-Burillo, S. 2006, ApJS, 164, 450
 \bibitem[2010]{xyz}
  Mart{\'i}n, S., Aladro, R., Mart{\'i}n-Pintado, \& Mauersberger, R. 2010, A\&A, 522, A62
 \bibitem[2011]{xyz}
  Mart\'{\i}n, S., Krips, M., Mart\'{\i}n-Pintado, J. et al. 2011, A\&A, 527, A36
 \bibitem[2012]{xyz}
  Mart\'{\i}n, S., Mart\'{\i}n-Pintado, J., Montero-Casta{\~n}o, Ho, P.~T.~P., \& Blundell, R. 2012, A\&A, 539, A29
 \bibitem[1996]{xyz}
  Mauersberger, R., Henkel, C., Wielebinski, R., Wiklind, T., \& Reuter, H.-P. 1996, A\&A, 305, 421
 \bibitem[2012]{xyz}
  Mikolaitis, {\^S}, Tautvai{\^s}ien{\'e}, G., Gratton, R., Bragaglia, A., \& Carretta, E. 2012, A\&A, 541, A137
 \bibitem[2005]{xyz}
  Milam, S.~N., Savage, C., Brewster, M.~A., Ziurys, L.~M., \& Wyckoff, S. 2005, ApJ, 634, 1126
 \bibitem[2005]{xyz}
  Mouhcine, M., Ferguson, H.~C., Rich, R.~M., Brown, T.~M., \& Smith, T.~E. 2005, ApJ 633, 810
 \bibitem[2006]{xyz}
  Muller, S., Gu{\'e}lin, M., Dumke, M., Lucas, R., \& Combes, F. 2006, A\&A, 458, 417
 \bibitem[2009]{xyz}
  Padovani, M., Walmsley, C.~M., Tafalla, M., Galli, D., \& M{\"u}ller, H.~S.~P. 2009, A\&A, 505, 1199
 \bibitem[2004]{xyz}
  Paglione, T.~A.~D., Yam, O., Tosaki, T., \& Jackson, J.~M. 2004, ApJ, 611, 835
 \bibitem[2011]{xyz}
  Papadopoulos, P.~P., Thi, W.-F., Miniati, F., \& Viti, S. 2011, MNRAS, 414, 1705
 \bibitem[1981]{xyz}
  Pence, W.~D. 1981, ApJ, 247, 473
 \bibitem[1996]{xyz}
  Peng, R., Zhou, S., Whiteoak, J.~B., Lo, K.~Y., \& Sutton, E.~C. 1996, ApJ, 470, 821
 \bibitem[1980]{xyz}
  Penzias, A.~A. 1980, Sci, 208, 663
 \bibitem[1991]{xyz}
  Puche, D., Carignan, C., \& van Gorkom, J.~H., 1991, AJ, 101, 456
 \bibitem[2005]{xyz}
  Rekola, R., Richer, M.~G., McCall, M.~L., et al. 2005, MNRAS, 361, 330
 \bibitem[2012]{xyz}
  Rimmer, P.~B., Herbst, E., Morata, O., \& Roueff, E. 2012, A\&A, 537, A7
 \bibitem[2012]{xyz}
  Riquelme, D., Amo-Baladr{\'o}n, M.~A., Mart\'{\i}n-Pintado, J. et al. 2010, A\&A 523, A51
 \bibitem[2011]{xyz}
  Sakamoto, K., Mao, R.-Q., Matsushita, S. et al. 2011, ApJ, 735, 19
 \bibitem[2007]{xyz}
  Sheffer, Y., Rogers, M., Federman, S.~R., Lambert, D.~L., \& Gredel, R. 2007, ApJ, 667, 1002
 \bibitem[1997]{xyz}
  Simon, R., Stutzki, J., Sternberg, A. \& Winnewisser, G. 1997, A\&A, 327, L9
 \bibitem[1983]{xyz}
  Skatrud, D.~D., De Lucia, F.~C., Blake, G.~A., \& Sastry, K.~V.~L.~N. 1983, J. Mol. Spectr. 99, 35
 \bibitem[2012]{xyz}
  Spielfiedel, A., Feautrier, N., Najar, F. et al. 2012, MNRAS, 421, 1891
 \bibitem[1989]{xyz}
  Stahl, O., Wilson, T.~L., Henkel, C., \& Appenzeller, I. 1989, A\&A, 221,321
 \bibitem[1998]{xyz}
  Taniguchi, Y. \& Ohyama, Y. 1998, ApJ, 507, L121
 \bibitem[1999]{xyz}
  Taniguchi, Y., Ohyama, Y., \& Sanders, D.~B. 1999, ApJ, 522, 214
 \bibitem[1968]{xyz}
  Thomson, R., Dalby, F.~W. 1968, Can. J. Phys. 46, 2815
 \bibitem[1985]{xyz}
  Turner, B.~E. 1985, ApJ, 299, 312
 \bibitem[1997]{xyz}
  Ulvestad, J.~S., \& Antonucci, R.~R.~J. 1997, ApJ, 488, 621
 \bibitem[2007]{xyz}
  van der Tak, F.~F.~S., Black, J.~H., Sch{\"o}ier, F.~L., Jansen, D.~J., \& van Dishoeck, E.~F. 2007, A\&A, 468, 627
 \bibitem[2004]{xyz}
  Wang, M., Henkel, C., Chin, Y.-N., et al. 2004, A\&A, 422, 883 
 \bibitem[2009]{xyz}
  Wang, M., Chin, Y.-N., Henkel, C., Whiteoak, J.~B., \& Cunnningham, M. 2009, ApJ, 690, 580
 \bibitem[1980]{xyz}
  Wannier, P.~G. 1980, ARA\&A, 18, 399
 \bibitem[1994]{xyz}
  Wilson, T.~L., \& Rood, R. 1994, ARA\&A, 32, 191
 \bibitem[1996]{xyz}
  Wouterloot, J.~G.~A., \& Brand, J. 1996, A\&AS, 119, 439
 \bibitem[2008]{xyz}
  Wouterloot, J.~G.~A., Henkel, C., Brand, J., \& Davis, G.~R. 2008, A\&A, 487, 237 

\end{thebibliography}
\end{document}